\documentclass[usenatbib]{mn2e}

\usepackage{color}
\usepackage{amsmath,amssymb}
\usepackage{graphicx}
\usepackage{aas_macros}
\graphicspath{ {maps/}, {stacked/}}

\usepackage{tablefootnote}

   \volume{{\rm in press}}
   
\title[Dense Gas Optical Depths in Nearby Galaxies]{Optical Depth Estimates and Effective Critical Densities of Dense Gas Tracers in the Inner Parts of Nearby Galaxy Discs}

\author[Mar\'ia J. Jim\'enez-Donaire et al.]{M. J. Jim\'enez-Donaire,$^{1}$\thanks{E-mail: m.jimenez@zah.uni-heidelberg.de}
F. Bigiel,$^{1}$
A. K. Leroy,$^{2}$
D. Cormier,$^{1}$
M. Gallagher,$^{2}$
\and
A. Usero,$^{3}$
A. Bolatto,$^{4}$
D. Colombo,$^{5}$
S. Garc\'ia-Burillo,$^{3}$
A. Hughes,$^{6,7}$
C. Kramer,$^{8}$
\and
M. R. Krumholz,$^{9}$
D. S. Meier,$^{10}$
E. Murphy,$^{11}$
J. Pety,$^{12,13}$
E. Rosolowsky,$^{14}$
\and
E. Schinnerer,$^{15}$
A. Schruba,$^{16}$
N. Tomi\v{c}i\'c,$^{15}$
and L. Zschaechner$^{15}$
\\
$^{1}$Institute f\"ur theoretische Astrophysik, Zentrum f\"ur Astronomie der Universit\"at Heidelberg, Albert-Ueberle Str. 2, 69120 Heidelberg, Germany.\\
$^{2}$Department of Astronomy, The Ohio State University, 140 W 18th Street, Columbus, OH 43210, USA\\
$^{3}$Observatorio Astron\'omico Nacional, Alfonso XII 3, E-28014 Madrid, Spain\\
$^{4}$Department of Astronomy and Laboratory for Millimeter-Wave Astronomy, University of Maryland, College Park, MD 20742, USA\\
$^{5}$Max-Planck-Institut f\"ur Radioastronomie, Auf dem H\"ugel 69, 53121, Bonn, 
Germany\\
$^{6}$CNRS, IRAP, 9 Av. colonel Roche, BP 44346, F-31028 Toulouse cedex 4, France\\
$^{7}$Universit\'{e} de Toulouse, UPS-OMP, IRAP, F-31028 Toulouse cedex 4, France\\
$^{8}$Instituto de Radioastronom\'ia Milim\'etrica
(IRAM), Av. Divina Pastora 7, N\'ucleo Central, E-18012 Granada, Spain\\
$^{9}$Research School of Astronomy \& Astrophysics, Australian National University, Canberra, ACT 2611, Australia\\
$^{10}$Department of Physics, New Mexico Institute of Mining and Technology, 801 Leroy Place, Soccoro, NM 87801, USA\\
$^{11}$National Radio Astronomy Observatory, 520 Edgemont Road, Charlottesville, VA 22903, USA\\
$^{12}$Institut de Radioastronomie Millim\'etrique (IRAM), 300 Rue de la Piscine, F-38406 Saint Martin d'H\`{e}res, France\\
$^{13}$Observatoire de Paris, 61 Avenue de l'Observatoire, F-75014 Paris, France\\
$^{14}$Department of Physics, University of Alberta, Edmonton, AB T6G 2E1, Canada\\
$^{15}$Max-Planck-Institut f\"ur Astronomie, K\"onigstuhl 17, D-69117 Heidelberg, Germany\\
$^{16}$Max-Planck-Institut f\"ur extraterrestrische Physik, Giessenbachstrasse 1, D-85748 Garching, Germany
}

\date{Accepted 2016 November 16. Received 2016 November 15; in original form 2016 August 26.
}

\pubyear{2016}

\begin{document}
\label{firstpage}
\pagerange{\pageref{firstpage}--\pageref{lastpage}}
\maketitle

\begin{abstract}

High critical density molecular lines like HCN~(1-0) or HCO$^+$~(1-0) represent our best tool to study currently star-forming, dense molecular gas at extragalactic distances. The optical depth of these lines is a key ingredient to estimate the effective density required to excite emission. However, constraints on this quantity are even scarcer in the literature than measurements of the high density tracers themselves. Here, we combine new observations of HCN, HCO$^+$ and HNC~(1-0) and their optically thin isotopologues H$^{13}$CN, H$^{13}$CO$^+$ and HN$^{13}$C~(1-0) to measure isotopologue line ratios. We use IRAM 30-m observations from the large program EMPIRE and new ALMA observations, which together target $6$ nearby star-forming galaxies. Using spectral stacking techniques, we calculate or place strong upper limits on the HCN/H$^{13}$CN, HCO$^+$/H$^{13}$CO$^+$ and HNC/HN$^{13}$C line ratios in the inner parts of these galaxies. Under simple assumptions, we use these to estimate the optical depths of HCN~(1-0) and HCO$^+$~(1-0) to be $\tau\sim$2-11 in the active, inner regions of our targets. The critical densities are consequently lowered to values between 5-20$\times 10^5$, 1-3$\times 10^5$ and 9$\times 10^4$ cm$^{-3}$ for HCN, HCO$^+$ and HNC, respectively. We study the impact of having different beam-filling factors, $\eta$, on these estimates and find that the effective critical densities decrease by a factor of $\frac{\eta_{12}}{\eta_{13}}\,\tau_{12}$. A comparison to existing work in NGC~5194 and NGC~253 shows HCN/H$^{13}$CN and HCO$^+$/H$^{13}$CO$^+$ ratios in agreement with our measurements within the uncertainties. The same is true for studies in other environments such as the Galactic Centre or nuclear regions of AGN-dominated nearby galaxies.

\end{abstract}






\section{Introduction}

Most studies of the molecular gas in nearby galaxies have focused on the lower-J carbon monoxide (CO) transitions, as these are the brightest molecular lines in other galaxies and are essential to characterise their total molecular masses and their dynamical properties \citep{2012ARA&A..50..531K}. However stars form out of physically small ($\sim 0.1-1$ pc) structures made primarily of dense molecular gas. Thus, CO spectroscopy alone is not sufficient to constrain the physical properties of the actual star-forming, molecular gas with densities $\sim \text{n}_{\text{H}_2} > 10^4$ cm$^{-3}$; observations of tracers with higher critical densities are necessary. 

These immediately star-forming structures remain too small to resolve in external galaxies. In particular, they are most easily studied using lines that preferentially select the dense, star-forming material. The low-J transitions HCN, HCO+, and HNC are among the brightest of these high critical density lines and can now be studied in other galaxies. These kind of observations are very challenging due to the faintness of these molecules and they have been mostly studied in regions of active star formation in the Milky Way \citep[e.g.][]{2003ARA&A..41...57L, 2005ApJ...635L.173W, 2010ApJ...723.1019H, 2010ApJ...724..687L, 2014prpl.conf...27A, 2016ApJ...824...29S}, including the Galactic Centre \citep[e.g.][]{1993ApJ...411..667M, 2005ApJ...622..346C, 2010A&A...523A..45R, 2013AAS...22134909B, 2015A&A...584A.102H}. Only a few nearby galaxies have been analysed using multi-line datasets of dense gas tracers. These are often individual pointings on bright galaxy centres \citep[e.g][]{2004ApJ...606..271G, 2006ApJ...640L.135G, 2008A&A...479..703G, 2008ApJ...681L..73B, 2008ApJ...677..262K, 2012A&A...539A...8G}, which have recently been expanded to cover the discs of star-forming galaxies \citep[e.g.][]{2013A&A...549A..17B,2015AJ....150..115U,2016ApJ...823...87S,2015ApJ...801...63M}.

Emission of common dense gas tracers like H$^{12}$CN, H$^{12}$CO$^+$ and HN$^{12}$C is often found to be optically thick. This leads to line trapping effects that allow the lines to be excited at lower densities than one would naively assume; at high optical depth, the critical density scales linearly with the inverse of the opacity for any optically thick line. That is, the density of gas traced by dense gas tracers depends on their optical depth. As a result, knowing the optical depth of lines like HCN~(1-0) and HCO$^+$~(1-0) is essential to estimating the masses and density structure of the dense interstellar medium (ISM).

A good observational way to test the opacity of dense gas tracers is to combine observations of the main dense gas tracing lines with those of their rarer isotopologues, e.g., those that include $^{13}$C instead of $^{12}$C. These lines, e.g., H$^{13}$CN and H$^{13}$CO$^+$, are usually optically thin because the molecules are less abundant by the isotopic ratio. Combined with few simplifying assumptions, the ratio of, e.g., H$^{13}$CN (1--0) to H$^{12}$CN (1--0) offers the prospect to test the opacity of dense gas tracers and so to help quantify the density to which these lines are sensitive. 

The obstacle to such measurements is the faintness of emission from dense gas tracer isotopologues. As a result, these have been detected only within few very nearby galaxies \citep{1998A&A...329..443H, 1998A&A...330..901C, 2004A&A...422..883W, 2013A&A...549A..39A}. In this paper, we use new data from the IRAM 30-m telescope and the Atacama Large Millimeter/submillimeter Array (ALMA) to make new measurements of these line ratios and use them to constrain the optical depth of dense gas tracers. We rely on two main data sets: whole disc mapping of NGC~5194 first presented in \citet{2016ApJ...822L..26B} (see Section \ref{sec:data}) and ALMA observations of four nearby star-forming galaxies: NGC~3351, NGC~3627, NGC~4254, NGC~4321 presented in detail by Gallagher, Leroy et al. (in prep.). We supplement these with observations of the nuclear starburst in NGC~253 \citep{2013Natur.499..450B}, see Section \ref{subsec:253}. All of these observations targeted both dense gas tracers and optically thin isotopologues. The properties of each target can be found in Table \ref{table:sample}

Beyond improving our general knowledge of dense gas tracers, a specific motivation of this study is to complement the IRAM large program EMPIRE ("EMIR Multiline Probe of the ISM Regulating Galaxy Evolution", Bigiel et al. in prep.). This survey, which first results are presented in \citet{2016ApJ...822L..26B}, is the first multi-line mapping survey that targets both dense gas (e.g., HCN, HCO$^+$, HNC) and total gas ($^{13}$CO, C$^{18}$O, $^{12}$CO) tracers across the whole area of a sample of nearby star-forming galaxies.

The paper is structured as follows. We describe the data in Section \ref{sec:data}, including the details of data reduction and uncertainty calculations. We present the methods used to analyse the spectra in Section \ref{sec:method} and the main results are shown in Section \ref{sec:results}, and are discussed in Section \ref{discussion}. Finally, we summarise our main conclusions in Section \ref{summary}. Throughout the paper, we will refer to the $^{12}$C isotopologues without special notation, e.g. HCN, etc., and mention explicitly if we refer to their isotopologues composed of other carbon isotopes.

\section{Observations and data reduction}
\label{sec:data}
\subsection{IRAM 30m Observations of NGC~5194}

NGC 5194 or M51 is a nearby spiral galaxy (SAbc) at a distance of 7.6 Mpc \citep{2002ApJ...577...31C}. It was mapped with the IRAM 30-m telescope at Pico Veleta over the course of 75 hours spread across seven runs during July and August 2012. The EMIR \citep{2012A&A...538A..89C} band E090 receiver was used in the dual polarisation mode to map the emission from several high density tracers. We reached a sensitivity value of RMS(T$_\text{mb}$)$=2.3\,$mK. The observations cover the whole active area of M51 and a frequency range of $\sim$86-106 GHz. This range covers HCN (1-0), HCO$^+$ (1-0), and HNC (1-0), as well as the analogous transitions from their rarer isotopologues H$^{13}$CN (1-0), H$^{13}$CO$^+$ (1-0), and HN$^{13}$C (1-0). More information about the observations, e.g. detailed frequency settings, integration time and system noise temperatures can be found in \citet{2016ApJ...822L..26B}.

All the scans were combined and exported using the GILDAS/CLASS software\footnote{http:/www.iram.fr/IRAMFR/GILDAS; for more information see \citet{2005sf2a.conf..721P}} which was also used for further analysis including linear baseline subtraction, smoothing, and flux measurements. The data were convolved to a common resolution of 28", which corresponds to 1.0 kpc linear resolution at the distance of M51. This means that our observations average together the emission from many giant molecular clouds (GMCs). The final velocity resolution after smoothing is 4 km s$^{-1}$. The uncertainties in the fluxes are estimated by means of the RMS per channel of width $dv$ within the signal-free parts of the spectra and multiplying this value by the square root of the number of channels covered by each line. We note that the receiver setup for the other 8 EMPIRE galaxies is slightly different to the one used for M51 and did not cover the $^{13}$C isotopologues.

\begin{table}
\caption{Properties of observed molecular lines}             
\label{table:1}      
\centering                          
\begin{tabular}{l c c c c}        

\hline\hline                 
 Name & Transition & Rest frequency & $E/k$ & $n_\textnormal{crit}$\\
  & & (GHz) & (K) & (cm$^{-3}$) \\
\hline                        
$^{13}$CO & 1-0 & 110.20 & 5.29 & 6.5$\times 10^2$\\
HCN & 1-0 & 88.63 & 4.25 & 5.0$\times 10^6$\\
HNC & 1-0 & 90.66 & 4.30 & 1.2$\times 10^6$\\
HCO$^{+}$ & 1-0 & 89.19 & 4.28 & 7.4$\times 10^5$\\
H$^{13}$CN & 1-0 & 86.34 & 4.14 & 9.7$\times 10^6$\\
H$^{13}$CO$^{+}$ & 1-0 & 86.75 & 4.16 & 6.7$\times 10^5$\\
HN$^{13}$C & 1-0 & 87.09 & 4.18 & 1.2$\times 10^6$\\
\hline                                   
\end{tabular}
 \\The critical densities were computed assuming optically thin transition lines for an excitation temperature of $20$K. The collision rates are adapted from the Leiden LAMDA database, \cite{2007A&A...468..627V}. For H$^{13}$CN and HN$^{13}$C the collisional coefficients for HCN and HNC from \citet{2010MNRAS.406.2488D} were used. For H$^{13}$CO$^+$ the collisional coefficients of HCO$^+$ from \citet{1999MNRAS.305..651F} were used.
\end{table}

\subsection{ALMA observations of NGC~3351, NGC~3627, NGC~4254, NGC~4321 and NGC~253}
\label{subsec:253}
We also study four galaxies, NGC~3351, NGC~3627, NGC~4254 and NGC~4321, which were observed during ALMA's Cycle 2 campaign (Gallagher, Leroy et al. in prep.). These observations covered the molecular lines listed in Table \ref{table:1} except for HN$^{13}$C. The data were processed using the ALMA pipeline and the CASA software \citep{2007ASPC..376..127M}, with details presented in Gallagher et al. (in prep.). After calibration, the data were continuum subtracted and then imaged into separate data cubes for each line. We reached a sensitivity value of RMS(T$_\text{mb}$)$=1.5\,$mK. A mild $u-v$ taper was applied to increase surface brightness sensitivity and then the data were convolved to have a $5\arcsec$ circular beam ($\sim$300 pc) and $10$~km~s$^{-1}$ wide channels.

We supplement these new observations with data for the nearby starburst galaxy NGC~253. These data, obtained as part of ALMA's Cycle 0 campaign, were originally presented by \citet{2015ApJ...801...63M} and \cite{2015ApJ...801...25L}. Like the new data above, these cover HCN and HCO$^+$, H$^{13}$CN, and H$^{13}$CO$^+$. The average beam size is $\sim$4" ($\sim$70 pc) over a field of view of approximately 1'.5 ($\sim$1.5 kpc). A summary of our data sets is given in Table \ref{table:1} and Table \ref{table:sample}.

\begin{table*}
\caption{\label{table:sample}Galaxy sample.}
\centering
\begin{tabular}{lcccccccccc}
\hline\hline
Source & RA & DEC & $i$ & PA & $r_{25}$ & $D$ & Morphology & $\Sigma_{\text{SFR}}$ & Inner aperture & Telescope\\
  & (2000.0) & (2000.0) & ($^o$) & ($^o$) & (') & (Mpc) & & ($\,M_\odot$ yr$^{-1}$ kpc$^{-2}$) & (kpc) & \\
\hline
NGC 3351 & 10:43:57.7 & 11:42:13.0 & 11.2 & 73 & 3.6 & 4.2 & SBb & 5.2 $\times 10^{-3}$ $^a$ 
& 0.20 & ALMA \\
NGC 3627 & 11:20:15.0 & 12:59:30.0 & 62 & 173 & 5.1 & 9.4 & SABb & 7.7$\times 10^{-3}$ $^a$ 
& 0.45 & ALMA \\
NGC 4254 & 12:18:50.0 & 14:24:59.0 & 32 & 55 & 2.5 & 14.4 & SAc & 18$\times 10^{-3}$ $^a$ & 0.70 & ALMA \\
NGC 4321 & 12:22:55.0 & 15:49:19.0 & 30 & 153 & 3.0 & 14.3 & SABbc & 9.0$\times 10^{-3}$ $^a$ & 0.69 & ALMA \\
NGC 5194 & 13:29:52.7 & 47:11:42.9 & 20 & 172 & 3.9 & 7.6 & Sbc & 20$\times 10^{-3}$ $^a$ 
& 1.84 & IRAM-30m \\
NGC 253 & 00:47:33.1 & -25:17:19.7 & 76 & 55 & 13.5 & 3.5 & SABc & 14$^b$ & 0.17 & ALMA \\
\hline
\end{tabular}
\\ $^a$ \citet{2013AJ....146...19L}, average SFR surface density inside 0.75$r_{25}$; $^b$ \citet{2015ApJ...801...25L}, taking SFR $\sim 1$ $\,M_\odot$ yr$^{-1}$ within $r_{50}$.
\end{table*}

\section{Methodology}
\label{sec:method}

Our main goal is to detect emission from the faint isotopologues, H$^{13}$CN, H$^{13}$CO$^{+}$ and HN$^{13}$C and compute the line ratios HCN/H$^{13}$CN, HCO$^+$/H$^{13}$CO$^+$ and HNC/HN$^{13}$C. The emission from these molecules is faint for individual lines of sight. To overcome this, we use spectral stacking. Because the line of sight velocity varies with the position in each galaxy due to rotation, we first re-grid velocity axis of each spectrum in order to set the local mean velocity of the bulk molecular medium to zero km/s. We then co-add the emission from many different lines of sight to produce an average, higher signal-to-noise spectrum. This procedure resembles the one described by \citet{2011AJ....142...37S} and \citet{2013AJ....146..150C}. We use $^{13}$CO emission to estimate the local mean velocity of the molecular gas that we use as a reference for the stacking. In NGC~5194, we use the $^{13}$CO map from PAWS \citep{2013ApJ...779...42S,2013ApJ...779...43P}. For the ALMA data, we use $^{13}$CO maps also observed by ALMA in a separate tuning.

We stack spectra across the inner parts of our target galaxies to maximise the signal-to-noise. We defined the active ``inner region'' of our targets by specifying a circular aperture with fixed angular size for the ALMA galaxies and different size for the M51 30m data. This leads to modestly varying physical aperture size from galaxy to galaxy. The size of the aperture for each galaxy is indicated in Figure \ref{densegas_maps} and noted in Table \ref{table:sample}. Specifically, for the ALMA galaxies we stack spectra inside a central 10" radius aperture. For M51 we picked a larger aperture of 50" in radius due to the lower resolution of the observations.

In order to measure spectral line parameters, we fit the stacked spectrum of each line with a single Gaussian. To perform the fit we use the MPFIT function in IDL. The free parameters we calculate from the fit are the line centre velocity, the peak intensity, and the velocity dispersion. We obtain the integrated HCN, H$^{13}$CN, HCO$^{+}$ and H$^{13}$CO$^{+}$ line intensities by integrating the fitted profile. We compute the uncertainties on the integrated intensity from the width of the profile and the rms noise measured from the signal-free part of the spectrum. If the peak intensity of the stacked spectrum is below 3$\sigma$, then we compute an upper limit to the integrated intensity. We take the limit to the integrated intensity to be the flux of a Gaussian profile with the FWHM of the corresponding $^{12}$C line and a peak three times the noise level for the $^{13}$C profile. This assumes that the $^{12}$C and $^{13}$C lines are well mixed, and, although the limit is formally stronger than $3\sigma$, this approach matches what one would identify as an upper limit by eye.

\section{Results}
\label{sec:results}

\begin{figure*}
\includegraphics[scale=0.85]{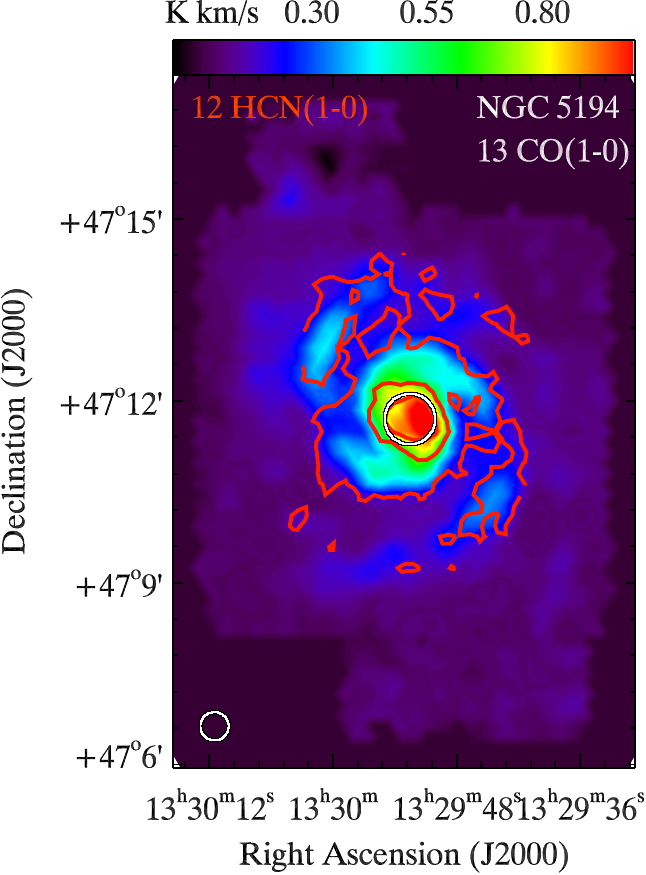}$\qquad$\,$\qquad$\,$\qquad$
\includegraphics[scale=0.8]{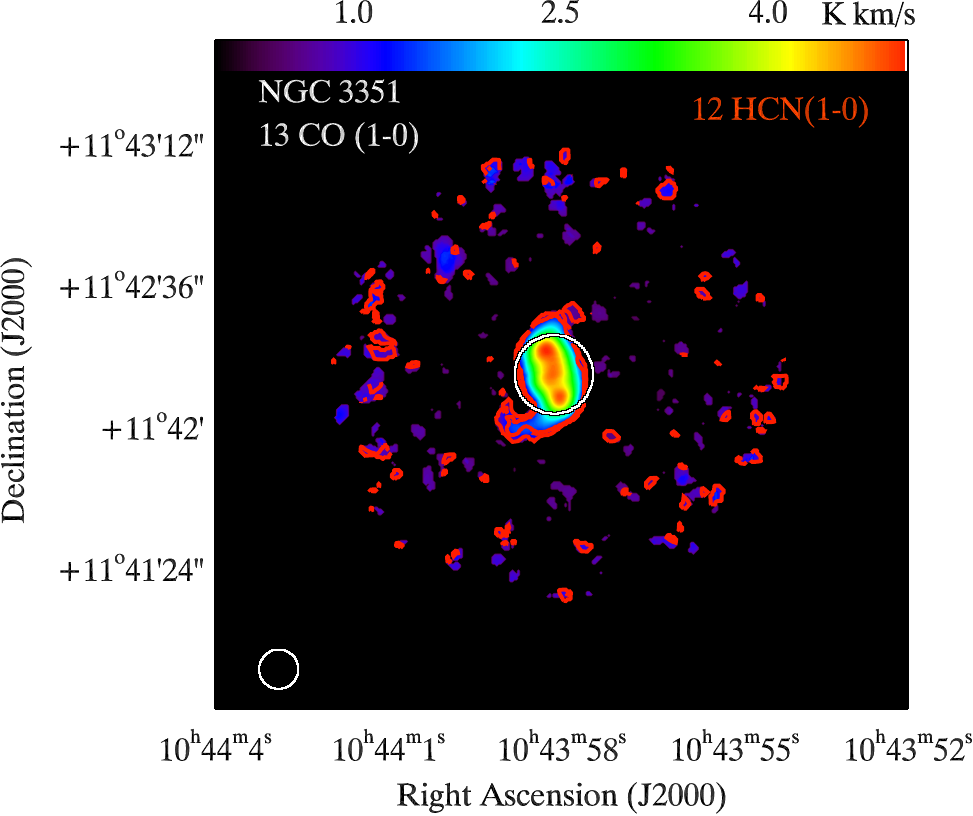}\\
\bigskip
\includegraphics[scale=0.8]{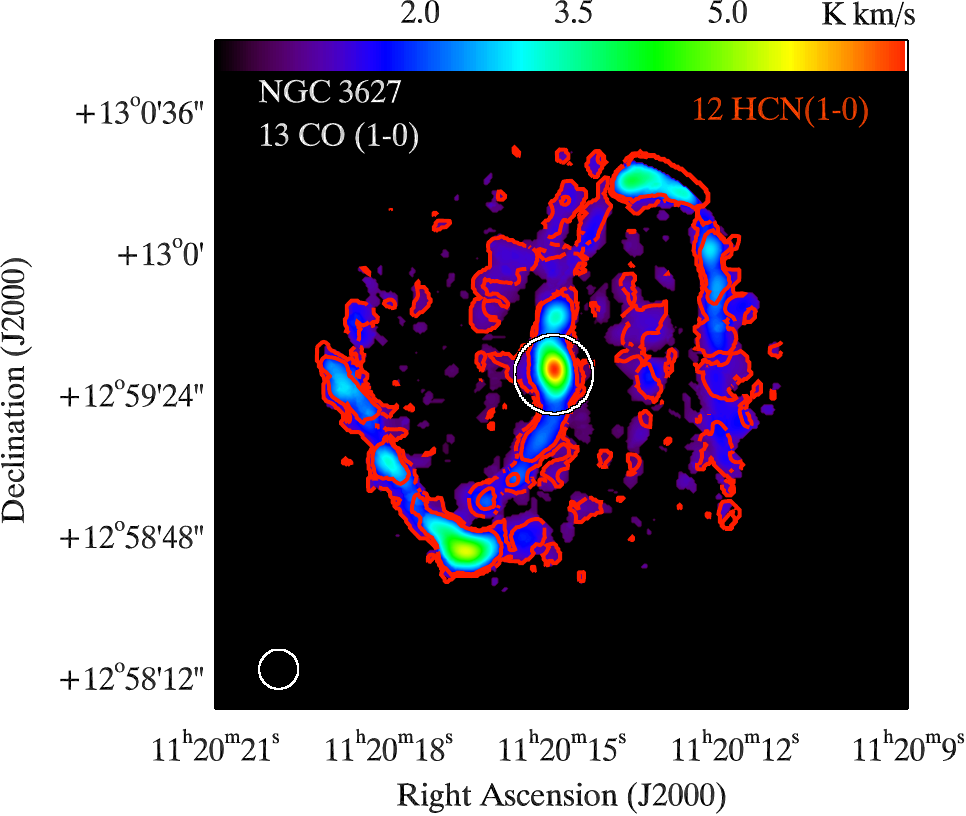}
\includegraphics[scale=0.8]{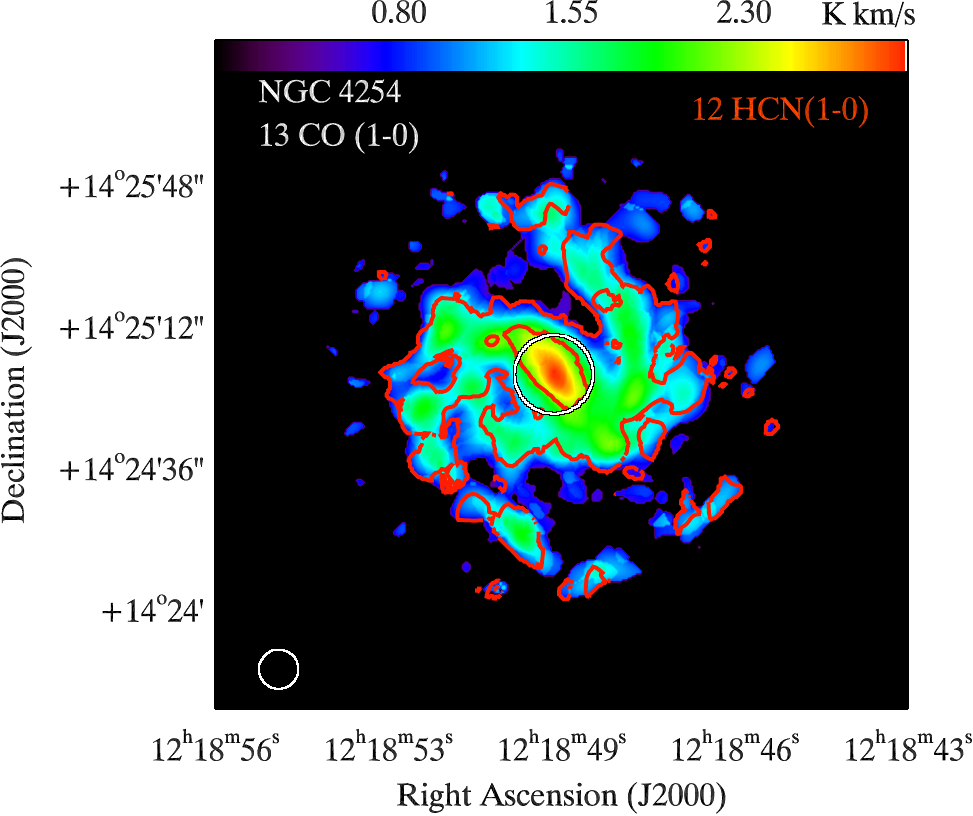}\\
\bigskip
\includegraphics[scale=0.8]{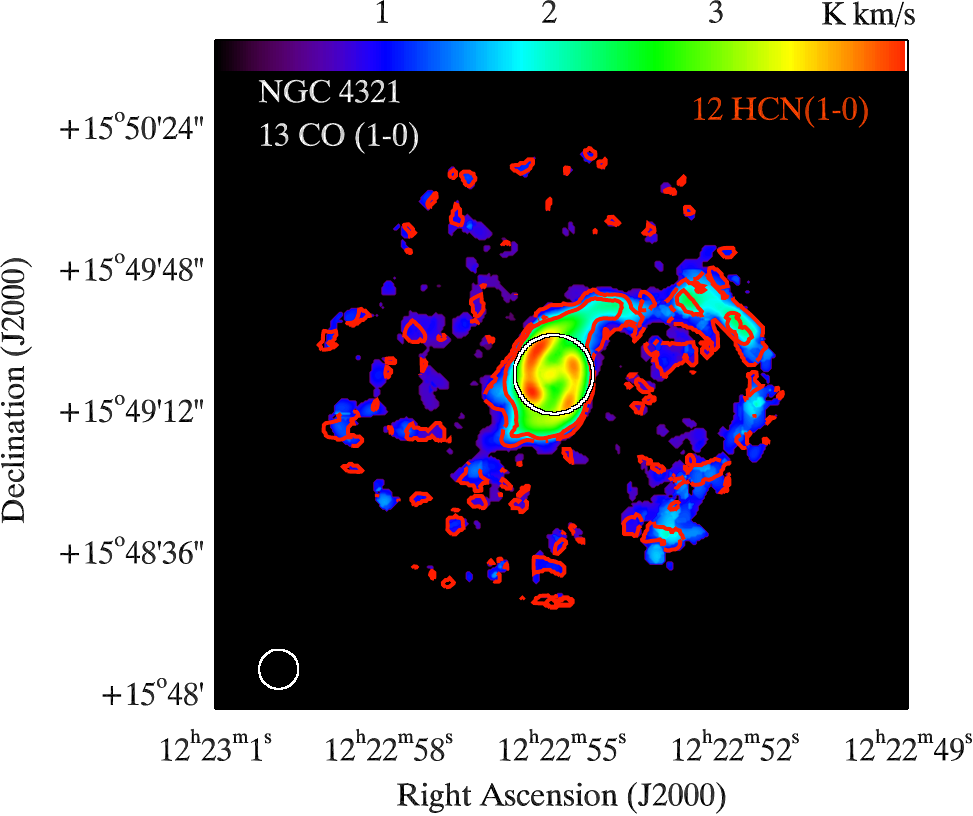}
\includegraphics[scale=0.8]{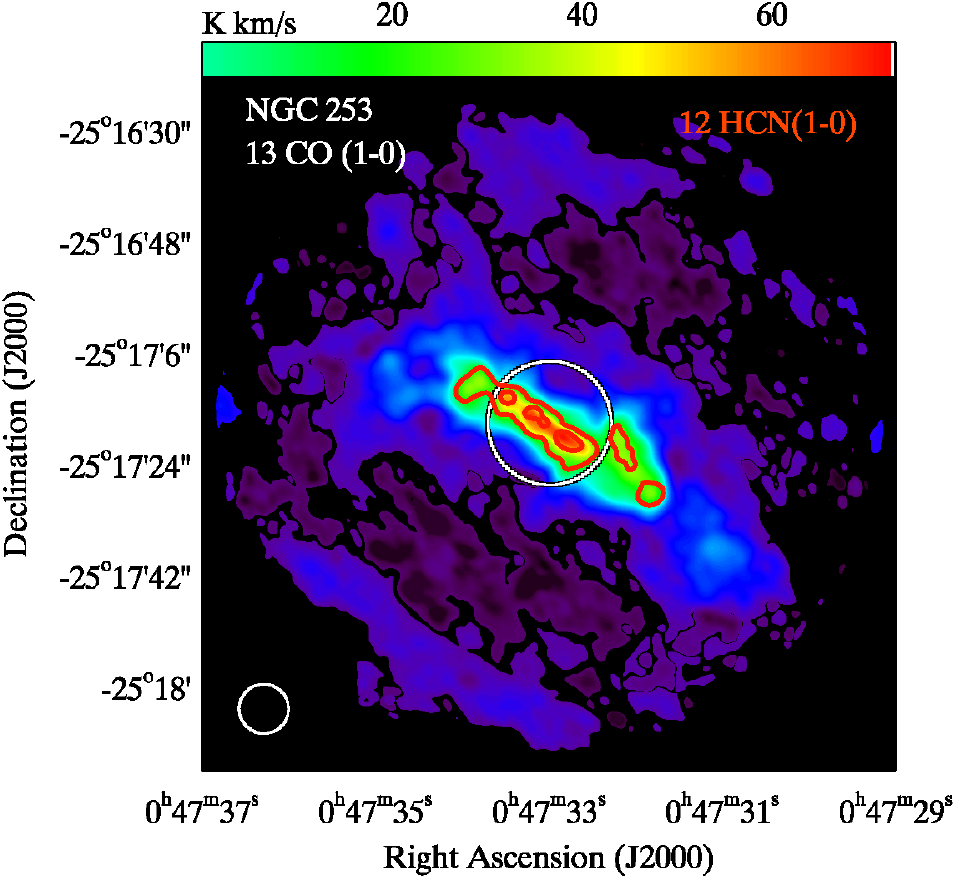}\\
\caption{$^{13}$CO (1-0) integrated intensity maps for our target galaxies. The red contours show HCN~(1-0) intensity levels between 3 and 40$\sigma$ for NGC\,5194, between 8 and 25$\sigma$ in NGC\,3351, NGC\,3627, NGC\,4254 and NGC\,4321; and between 30 and 60$\sigma$ for NGC\,253. The white circle in the inner part of each target shows the selected region for each galaxy to stack the spectra, see details in Section \ref{sec:results}.}
\label{densegas_maps}
\end{figure*}

Our targets, listed in Table \ref{table:sample}, are all massive, star-forming, nearby disc galaxies. These are among the closest and hence best-studied star-forming galaxies. Our goal is to measure the ratios among dense gas tracers and their (presumably) optically thin $^{13}$C isotopologues. Figure \ref{densegas_maps} shows the distribution of $^{13}$CO~(1-0) and HCN~(1-0) emission from each targets. We use the $^{13}$CO emission, shown in colour, to indicate the distribution of the total molecular gas reservoir. It also serves as the reference for our spectral stacking. Bright $^{13}$CO emission is widespread across our targets, and we see it at good signal to noise everywhere that we see HCN. Red contours show the location of significant HCN (1-0) emission, tracing higher density gas. Although detected at lower significance than the $^{13}$CO, HCN is also widespread across our targets. In both lines and all galaxies, the brightest emission comes from the central regions of the target (the NGC\,253 data cover only the centre), with emission which extends along the spiral arms.

Although we detect dense gas tracers over large areas in each target, in most of our targets only the inner, bright regions yield useful stacked constraints on the isotopologue line ratios. We experimented with also stacking over the discs of our targets, but the emission of the $^{12}$C lines over this region already has only modest signal-to-noise. Stacking over the extended discs of galaxies typically yielded upper limits that barely constrained the $^{13}$C lines to be fainter than the $^{12}$C lines and so offered little constraint on the optical depth. Thus, we concentrate mainly on the bright inner regions for the remainder of this paper. As Figure \ref{densegas_maps} shows, these are bright in $^{13}$CO, HCN, and HCO$^+$. Here we detect the $^{13}$C lines in some cases and in all cases place limits comparable to typical $^{12}$CO/$^{13}$CO ratios in nearby galaxies (though see discussion below).

\begin{figure*}   
\begin{center}
\includegraphics[scale=0.45]{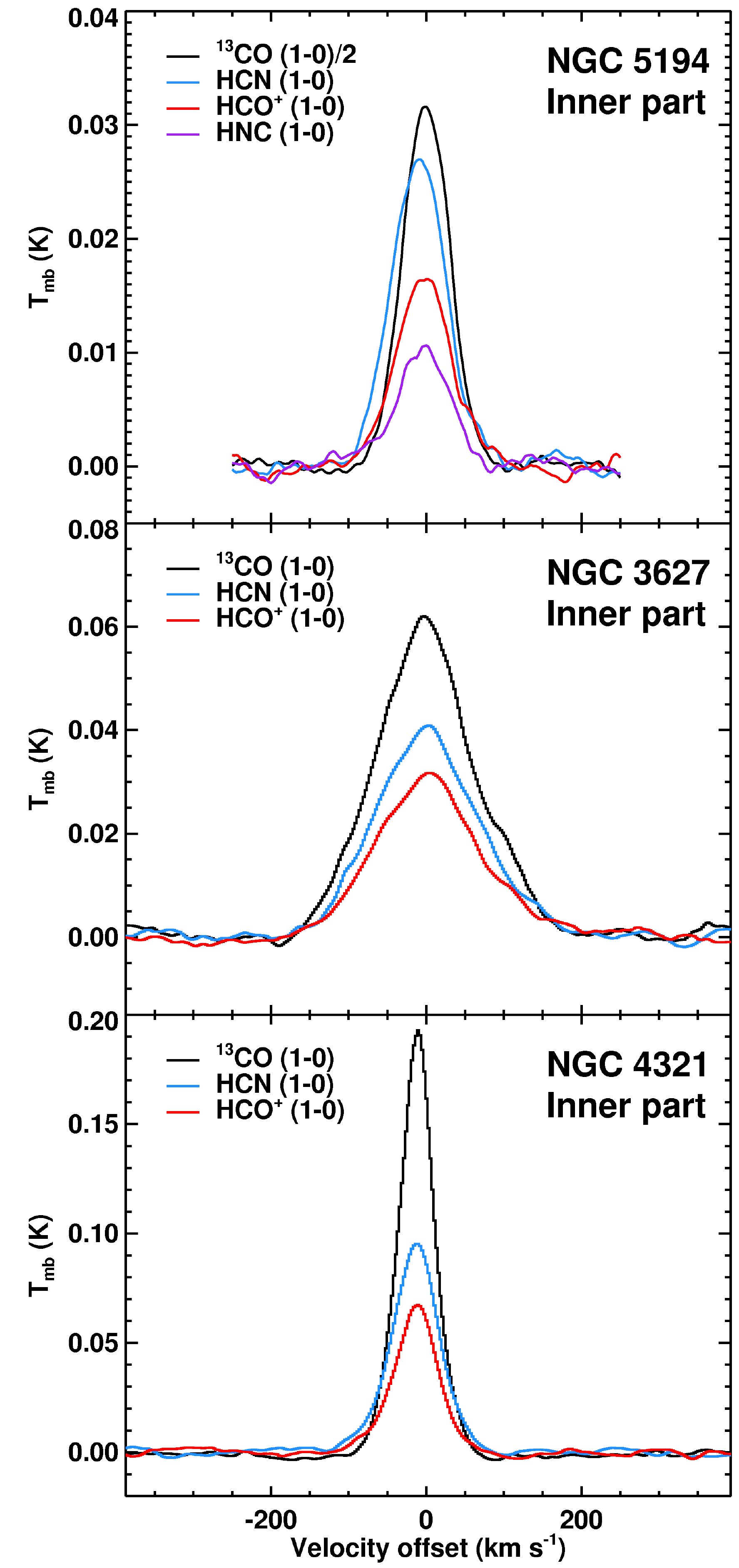}
\includegraphics[scale=0.45]{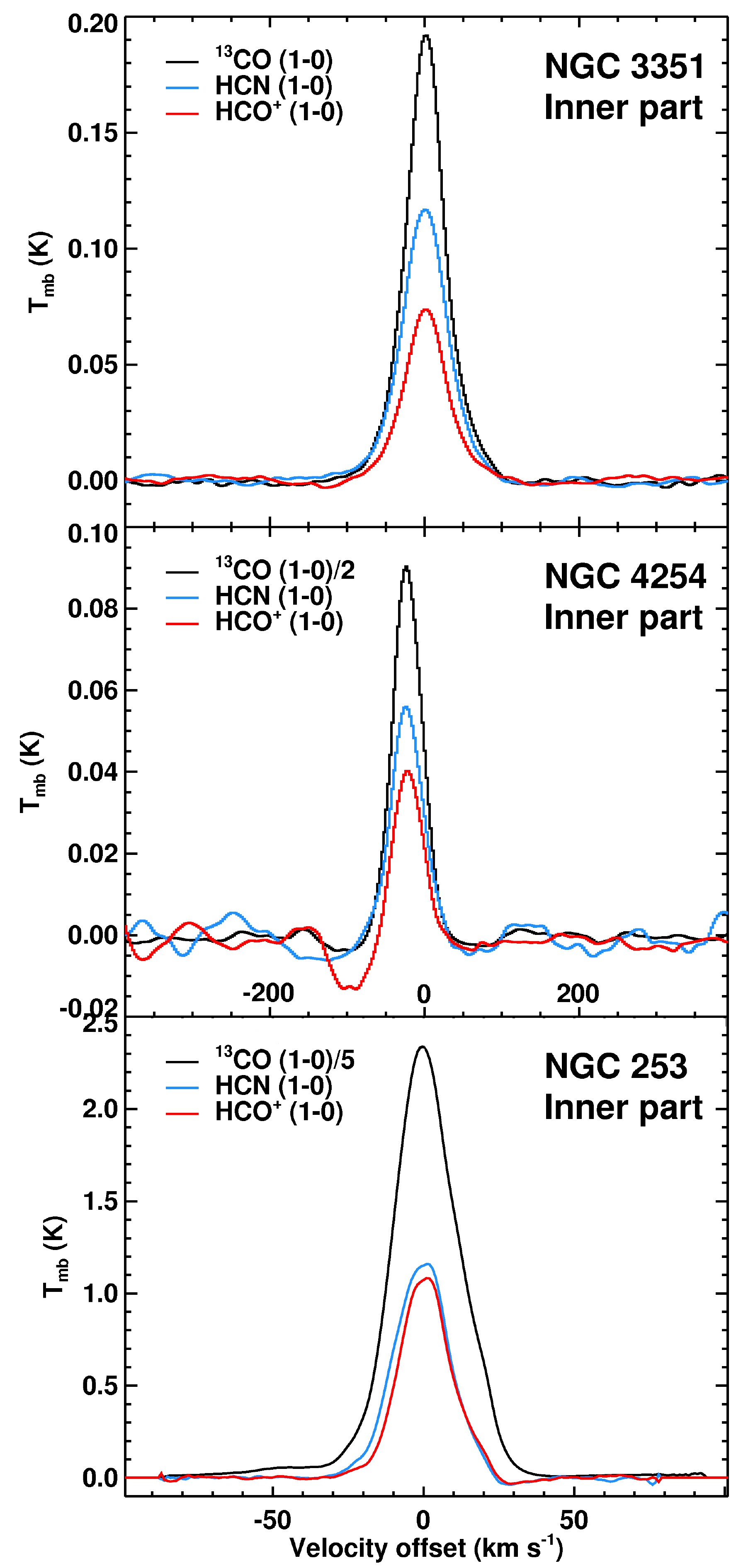}
\end{center}
\caption{Stacked $^{13}$CO, HCN, and HCO$^+$ spectra for the bright inner regions (radius $10\arcsec$) in our target galaxies. We also include the stacked HNC emission for NGC~5194. $^{13}$CO, which serves as our reference for the stacking, is scaled to match the intensity scale of the other lines. The $^{13}$CO and dense gas tracers are all detected at very high signal-to-noise. The stacked dense gas tracers show good agreement with the mean $^{13}$CO velocity and all lines show similar line widths, indicative of being well mixed on the scale of the beams ($\sim$ few hundred pc to kpc) for our data. Table \ref{table:lines} reports Gaussian fits to the lines.}
\label{centers}
\end{figure*}

Figures \ref{centers} and \ref{isotopologues} show the stacked spectra for $^{13}$CO, the HCN, HCO$^+$, HNC lines, and their $^{13}$C isotopologues for the bright, inner region of each target galaxy. Table \ref{table:lines} lists the line parameters derived from a Gaussian fit. Table \ref{table:ratios} reports key ratios derived from comparing these fitted parameters.

Figure \ref{centers} shows that we detect $^{13}$CO, HCN, HCO$^+$, and (in NGC 5194) HNC at very high significance over these active regions. The ordering of intensity is the same across all systems, with the intensity of $^{13}$CO $>$ HCN $>$ HCO$^+$. In NGC 5194, where we also measure HNC, this line is the weakest of the four.

Figure \ref{isotopologues} shows the stacked spectra for the $^{13}$C isotopologues of the dense gas tracers. Here, scaled versions of the $^{12}$C lines appear for reference. The $^{13}$C isotopologues of the dense gas tracers are all far fainter than the $^{12}$C lines. We do securely detect H$^{13}$CN in NGC~3351, NGC~3627, and NGC~253. In the other targets, we place firm upper limits on the intensity of these lines. We detect H$^{13}$CO$^+$ only in NGC~253, placing limits in the other systems.

\begin{figure*}
\begin{center}
\includegraphics[scale=0.45]{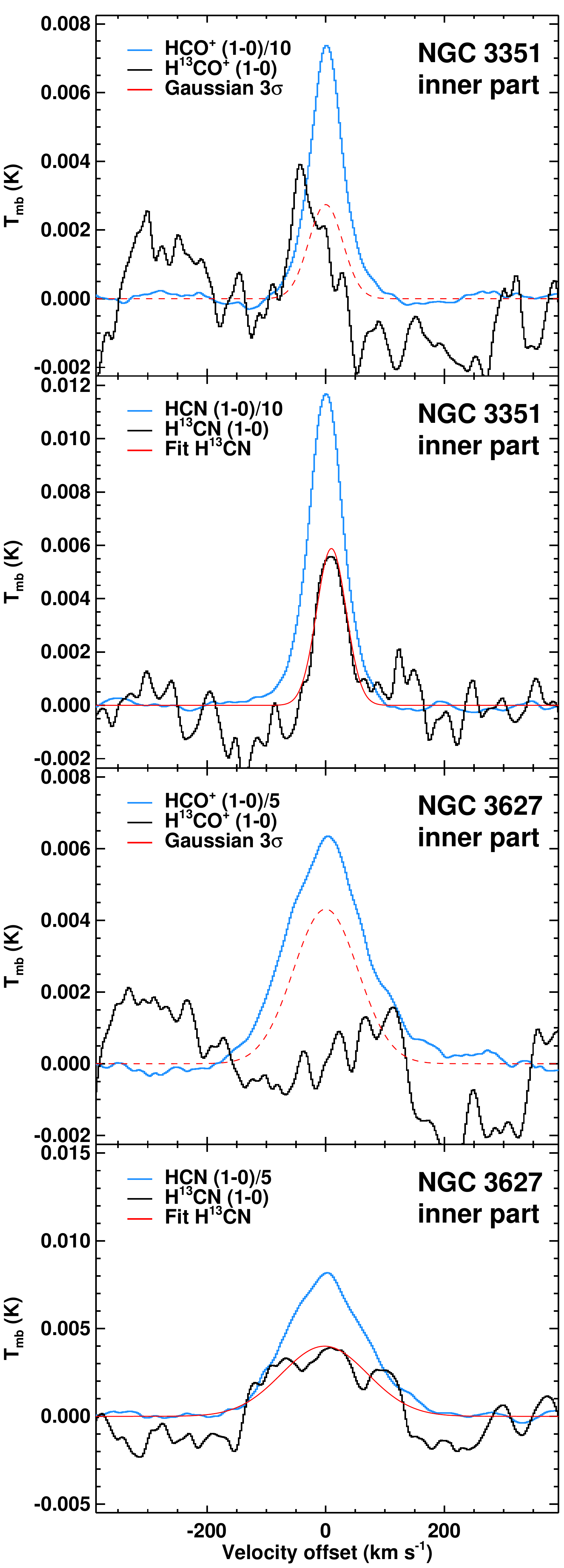}
\includegraphics[scale=0.45]{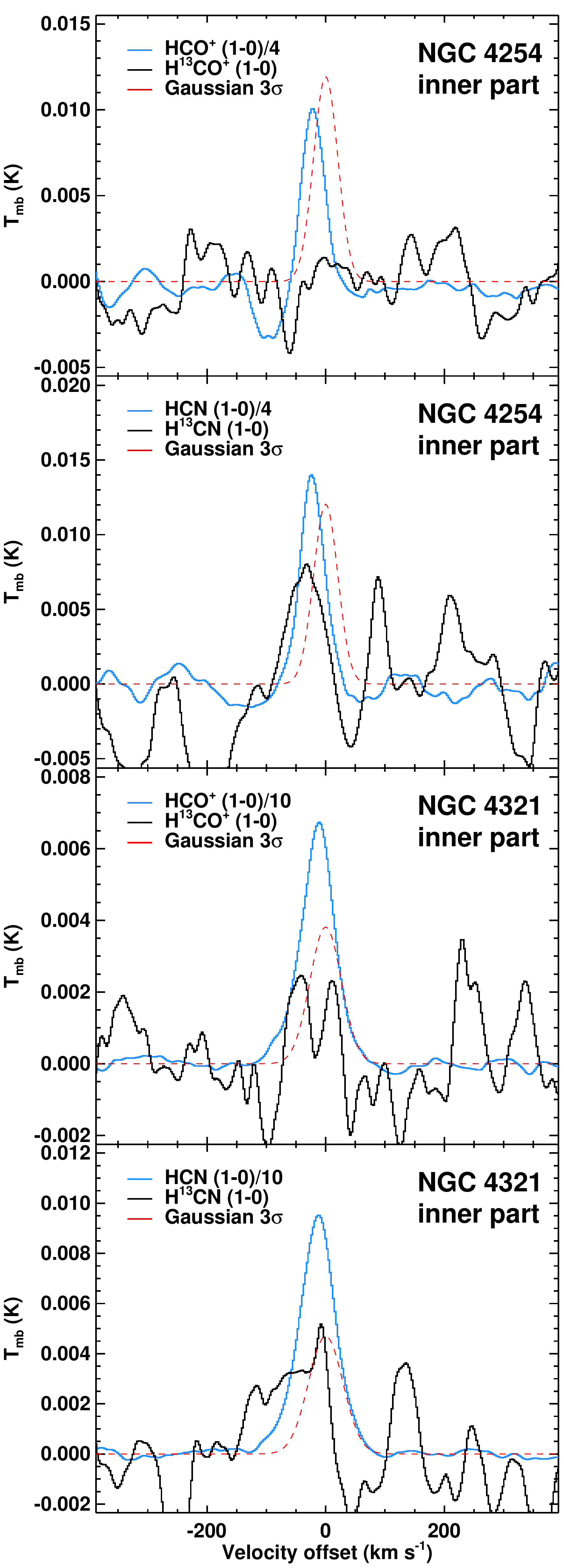}
\end{center}
\caption{Stacked H$^{13}$CN and H$^{13}$CO$^+$ spectra for the bright inner regions of our target galaxies. We show scaled versions of the stacked HCN and HCO$^{+}$ spectra from Figure \ref{centers} for reference. These  spectra show the expectation for a $^{12}$C-to-$^{13}$C line ratios of 1-to-5 to 1-to-10, common values for the $^{12}$CO-to-$^{13}$CO ratio in galaxy discs. The red line indicates our working upper limit (dashed) or fit (solid) to the iostopologue spectrum. Limits and fit parameters are reported in Table \ref{table:lines}.}
\label{isotopologues}
\end{figure*}

\begin{figure*}   
\begin{center}
\includegraphics[scale=0.45]{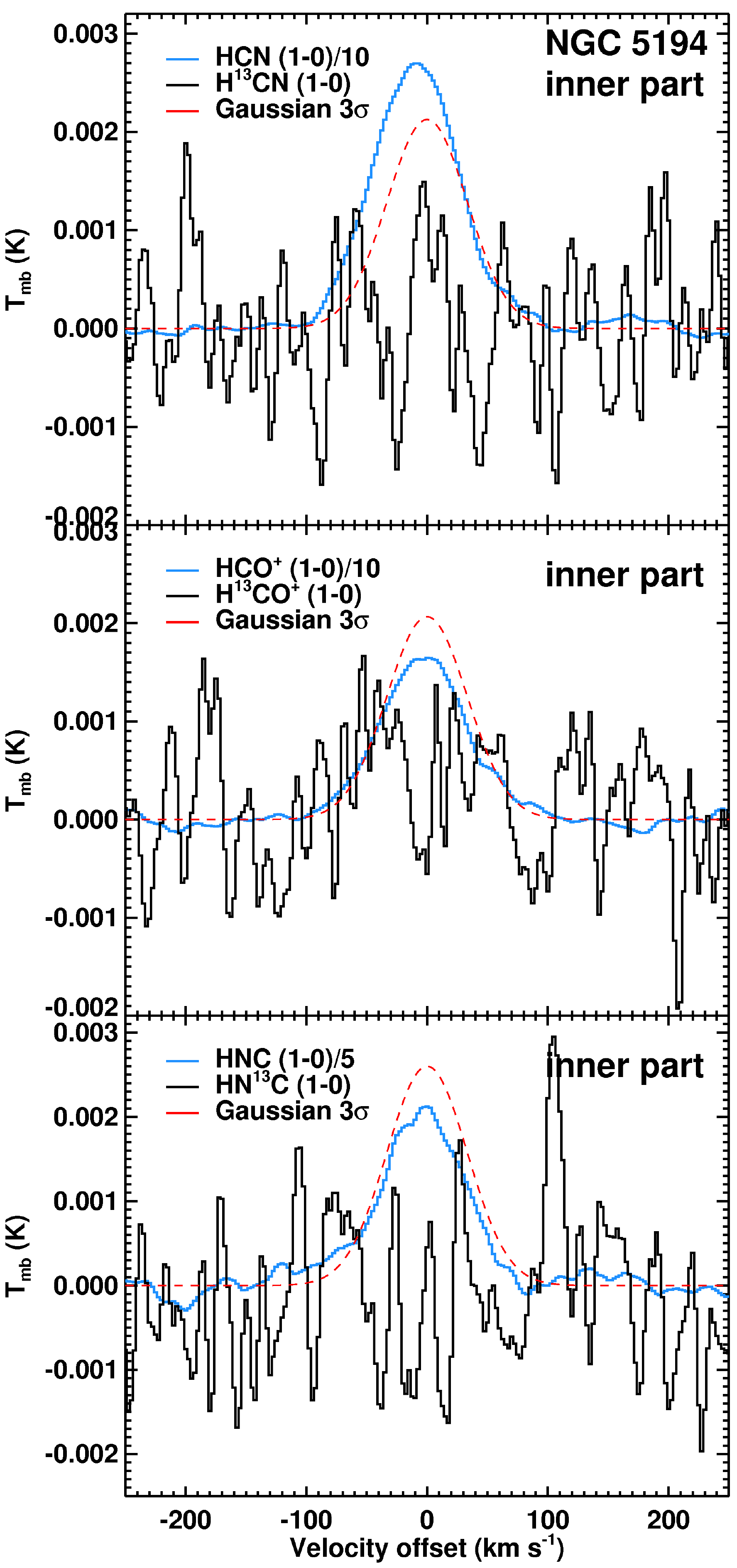}
\includegraphics[scale=0.45]{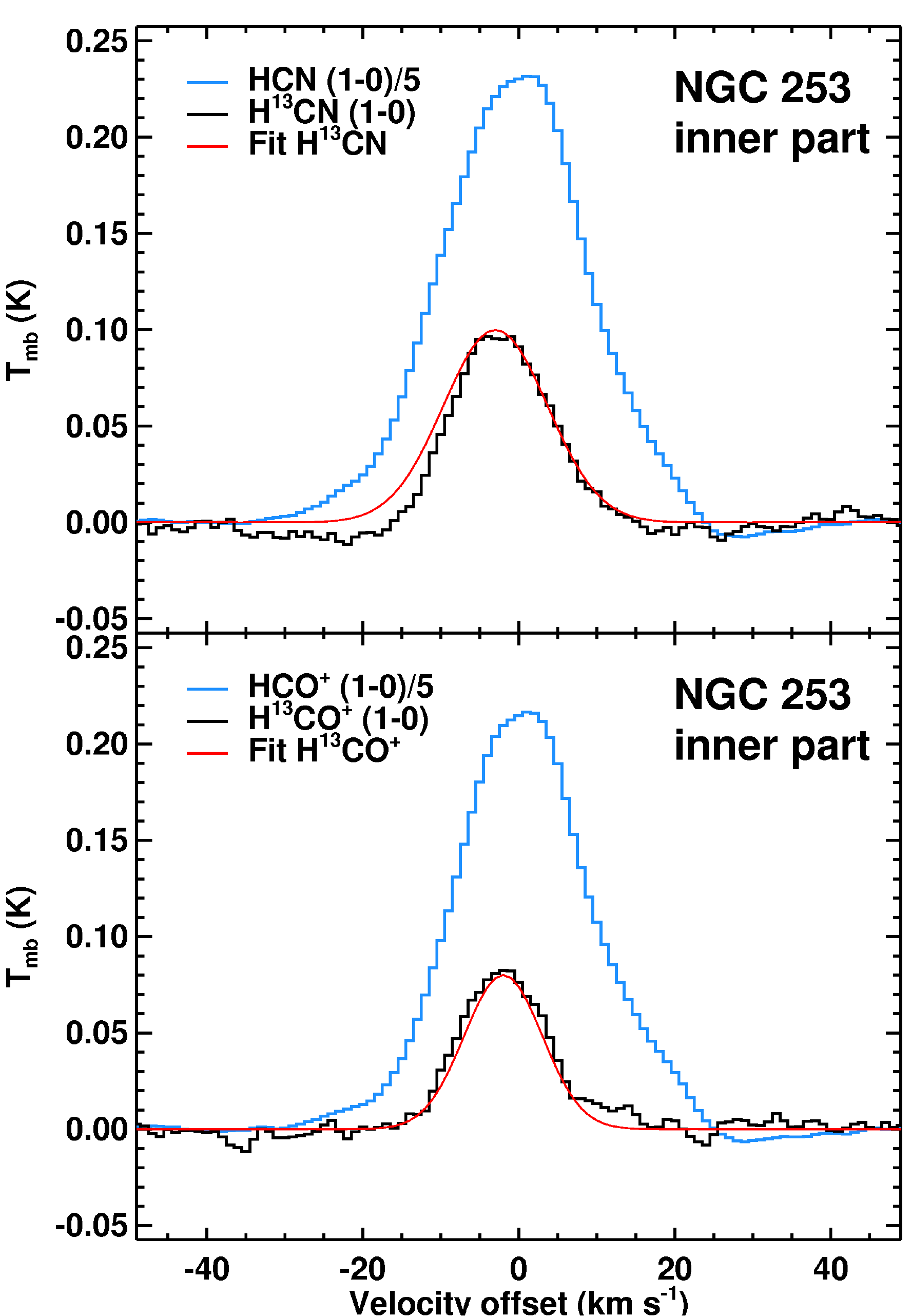}
\end{center}
\contcaption{}
\end{figure*}

\begin{table*}
\caption{Spectral line parameters for the galaxies derived from the stacked spectra in Figures 2 and 3.}
\label{table:lines}      
\centering          
\begin{tabular}{l l c c c c}        
\hline\hline                 

Galaxy & Molecule & Integrated intensity & Line width & Peak & rms\\
 & & (K km s$^{-1}$) & (km s$^{-1}$) & $\times 10^{-2}$(K) & (mK)\\
 \hline
NGC~3351 & $^{13}$CO& 12.0$\pm$0.1 & 62$\pm$10 & 19.4$\pm$0.4 & 1.1\\
 & HCN& 8.4$\pm$0.1 & 67$\pm$11 & 11.8$\pm$0.6 & 1.5\\
 & H$^{13}$CN & 0.4$\pm$0.1 & 59$\pm$10 & 0.7$\pm$0.1 & 0.9\\
 & HCO$^+$ & 5.0$\pm$0.1 & 64$\pm$11 & 7.4$\pm$0.2 & 1.4\\
 & H$^{13}$CO$^+$ & $<$0.3 & --  & $<$0.4 & 1.3\\
\hline
NGC~3627 & $^{13}$CO & 7.4$\pm$0.1 & 121$\pm$14 & 6.8$\pm$0.4 & 1.0\\
 & HCN & 5.1$\pm$0.1 & 129$\pm$14 & 4.5$\pm$0.1 & 0.8\\
 & H$^{13}$CN & 0.7$\pm$0.1 & 170$\pm$17 & 0.4$\pm$0.1 & 1.0\\
 & HCO$^+$ & 4.9$\pm$0.1 & 128$\pm$14 & 3.5$\pm$0.1 & 1.0\\
 & H$^{13}$CO$^+$ & $<$0.4 & -- & $<$0.5 & 1.4\\
\hline
NGC~4254 & $^{13}$CO & 8.5$\pm$0.2 & 44$\pm$9 & 18.1$\pm$0.4 & 2.3\\
 & HCN & 2.9$\pm$0.2 & 48$\pm$9 & 5.5$\pm$0.2 & 3.3\\
 & H$^{13}$CN & $<$0.5 & -- & $<$1.3 & 4.0\\
 & HCO$^+$ & 2.0$\pm$0.1 & 44$\pm$9 & 3.8$\pm$0.1 & 1.8\\
 & H$^{13}$CO$^+$ & $<$0.3 & -- & $<$1.0 & 1.9\\
\hline
NGC~4321 & $^{13}$CO & 10.7$\pm$0.1 & 55$\pm$10 & 19.3$\pm$0.4 & 1.1\\
 & HCN & 6.9$\pm$0.1 & 69$\pm$11 & 9.7$\pm$0.2 & 1.4\\
 & H$^{13}$CN & $<$0.3 & -- & $<$1.0 & 1.6\\
 & HCO$^+$ & 4.6$\pm$0.1 & 63$\pm$11 & 6.8$\pm$0.2 & 1.4\\
 & H$^{13}$CO$^+$ & $<$0.3 & -- & $<$1.0 & 1.3\\
 \hline
 NGC~5194 & $^{13}$CO & 4.6$\pm$0.1 & 71$\pm$6 & 6.4$\pm$0.2 & 0.9\\
 & HCN & 2.2$\pm$0.1 & 76$\pm$8 & 2.7$\pm$0.1 & 0.6\\
 & H$^{13}$CN & $<$ 0.2 & -- & $<$0.2 & 0.7\\
 & HCO$^+$ & 1.4$\pm$0.1 & 78$\pm$10 & 1.6$\pm$0.3 & 0.6\\
 & H$^{13}$CO$^+$ & $<$ 0.2 & -- & $<$0.2 & 0.7\\
 & HNC & 0.8$\pm$0.1 & 77$\pm$8 & 1.1$\pm$0.2 & 0.8\\
 & HN$^{13}$C & $<$ 0.2 & -- & $<$0.2 & 0.9\\
\hline
NGC~253 & $^{13}$CO & 312$\pm$9 & 25$\pm$6 & 1250$\pm$40 & 4.0\\
 & HCN & 32$\pm$2 & 22$\pm$4 & 120$\pm$4 & 9.0\\
 & H$^{13}$CN & 1.9$\pm$0.4 & 19$\pm$5 & 9.0$\pm$0.3 & 5.6\\
 & HCO$^+$ & 29$\pm$3 & 21$\pm$5 & 109$\pm$4 & 9.2\\
 & H$^{13}$CO$^+$ & 1.2$\pm$0.4 & 18$\pm$4 & 6.1$\pm$0.2 & 8.0\\

\hline
\end{tabular}
\end{table*}

\begin{table*}
\caption{Computed line ratios for the galaxies.}
\label{table:ratios}      
\centering          
\begin{tabular}{l r r r r r r}        
\hline\hline                 
Galaxy &NGC 3351& NGC 3627& NGC 4254 & NGC 4321 &NGC 5194 & NGC 253\\
\hline                        

HCN/H$^{13}$CN & 21$\pm$3 & 7$\pm$1 & $>$6 & $>$23 & $>$11 & 17$\pm$1\\

HCO$^+$/H$^{13}$CO$^+$ & $>$17 & $>$12 & $>$7 & $>$15 & $>$7 & 24$\pm$2\\

HNC/HN$^{13}$C & -- & -- & -- & -- & $>$4 & --\\

HCN/HCO$^+$ & 1.7$\pm$0.1 & 1.0$\pm$0.1 & 1.5$\pm$0.1 & 1.5$\pm$0.1 & 1.6$\pm$0.1  & 1.1$\pm$0.1 \\

$^{13}$CO/HCN & 1.4$\pm$0.2 & 1.5$\pm$0.1 & 3.0$\pm$0.3 & 1.6$\pm$0.2 & 2.1$\pm$0.1 & 10$\pm$1 \\

\hline                  
\end{tabular}
\end{table*}

\subsection{Line Ratios}
\label{section:ratios}

\begin{table}
\caption{Integrated intensity ratios from the literature.}             
\label{table:previous}      
\centering                          
\begin{tabular}{l c c c}        
\hline\hline                 
 Target & HCN/H$^{13}$CN & HCO$^{+}$/H$^{13}$CO$^{+}$ & Ref.\\
\hline                        
NGC 1068 & $\sim$16 & $\sim$20 & 1\\
Orion B & -- & $\sim$25 & 2\\
Sgr A & -- & $\sim$15 & 3\\
Galactic Centre & $\sim$11 & $\sim$21 & 4\\
NGC 5194 & $\sim$25 & $\sim$30 & 5\\
NGC 253 & $\sim$15 & $\sim$19 & 6\\
\hline                                   
\end{tabular}
\\(1) \citet{27050d31659a4cb1bdc9a59f0dde3f51}; (2) J. Pety (priv. comm.); (3) \citet{2010A&A...523A..45R}; (4) \citet{2015A&A...584A.102H}; (5) \citet{2014ApJ...788....4W}; (6) \citet{2015ApJ...801...63M}.
\end{table}

\begin{figure*}
\includegraphics[scale=0.5]{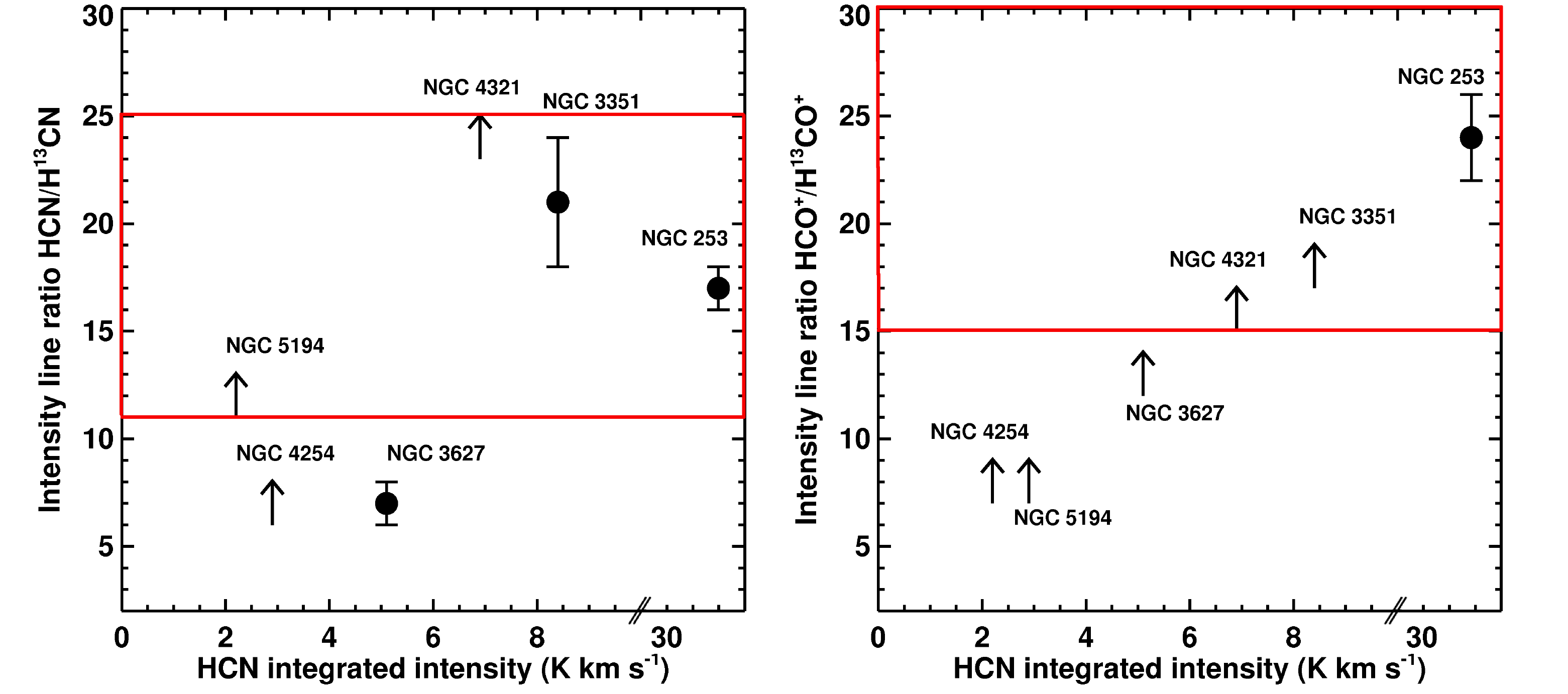}
\caption{Comparison between the line ratios measured in our sample of galaxies and the literature values, summarised in Table \ref{table:previous}. Circles show measurements from the stacked spectra in the inner regions of galaxies, whereas the arrows indicate upper limits. Note the non-continuous x-axis, as the HCN integrated intensity for NGC\,253 is much higher than for the other galaxies. The red boxes show the regime of values found for the sources in the literature in Table \ref{table:previous}, agreeing largely with the values and upper limits we derive.}
\label{comparison}
\end{figure*}

Table \ref{table:ratios} reports the line ratios averaged over the active regions, which are the main objective of our investigations. We measure HCN/H$^{13}$CN integrated intensity ratios ranging from $7-21$ (see Table \ref{table:ratios} for more details) and we constrain the ratio to be $\gtrsim 6$ for the entire sample. For HCO$^+$/H$^{13}$CO$^+$, we constrain the ratio to be $\gtrsim 7$ in our sample, and we measure a ratio of $24$ in NGC 253. For comparison, $^{12}$CO/$^{13}$CO ratios of $\sim 6$--$10$ are typically measured in the discs of galaxies, including the Milky Way and M51. The $^{12}$C/$^{13}$C ratios for the dense gas tracers are significantly higher than those for CO in our targets. Possible explanations are: lower optical depth in the dense gas tracers, isotopic abundance variations in the active parts of galaxies, or different filling factors for the $^{12}$C and $^{13}$C lines. We will discuss this further in Section \ref{discussion}

How do our measured ratios compare to previous measurements? Table \ref{table:previous} summarises literature measurements of isotopologue line ratios from the Milky Way and nearby galaxies, including two that overlap our targets (NGC 253 and NGC 5194). Broadly, these agree with our measurements, with ratios often $\gtrsim 10$ and sometimes $\gtrsim 20$. In detail, \citet{2014ApJ...788....4W} measured line ratios in NGC\,5194 and found HCN/H$^{13}$CN = 27$\pm$18, HCO$^+$/H$^{13}$CO$^+$ = 34$\pm$29, and HNC/HN$^{13}$C $>$ 16, consistent with our results for that system. Using the same data that we employ for NGC~253, \citet{2015ApJ...801...63M} found similar ratios of HCN/H$^{13}$CN, and HCO$^+$/H$^{13}$CO$^+$ \citep[see][]{2015ApJ...801...25L}. Observations of NGC\,1068 by \citet{27050d31659a4cb1bdc9a59f0dde3f51} found HCO$^+$/H$^{13}$CO$^+$ $=20\pm1$, HCN/H$^{13}$CN $=16\pm1$ and HNC/HN$^{13}$C $=38\pm6$. This galaxy is characterised by a strong AGN, which should influence the surrounding chemistry. Two of our targets also host AGN, NGC~5194 and NGC~3627, however we do not see any significant difference in the measured line ratios compared to the rest of the sample. Though we note that at the coarse resolution of our observations such effects may be difficult to isolate.

Studies focused on Milky Way yield a similar picture. \citet{2010A&A...523A..45R} carried out a large-scale survey of the Galactic Centre region and found HCO$^+$/H$^{13}$CO$^+$ $\sim$ 15 in Sgr A, similar to those found here. \citet{2015A&A...584A.102H} observed the Circumnuclear Disc of the Galactic Centre and reported ratios of HCO$^+$/H$^{13}$CO$^+$ $\sim$ 21 and HCN/H$^{13}$CN $\sim$ 11. In the Solar Neighbourhood, J. Pety (priv. comm.) found HCO$^+$/H$^{13}$CO$^+$ $\sim$ 25 in Orion B.

Figure \ref{comparison} plots our measured ratios along with these literature values. Red boxes highlight an indicative range of values for each ratio: $\sim 15$--$25$ for HCN/H$^{13}$CN and $\sim 15$--$30$ for HCO$^+$/H$^{13}$CO$^+$. We discuss the interpretation of these ratios in Section \ref{discussion}.

\subsection{Line Widths}
\label{section:widths}

\begin{figure}
\begin{center}
\includegraphics[scale=0.45]{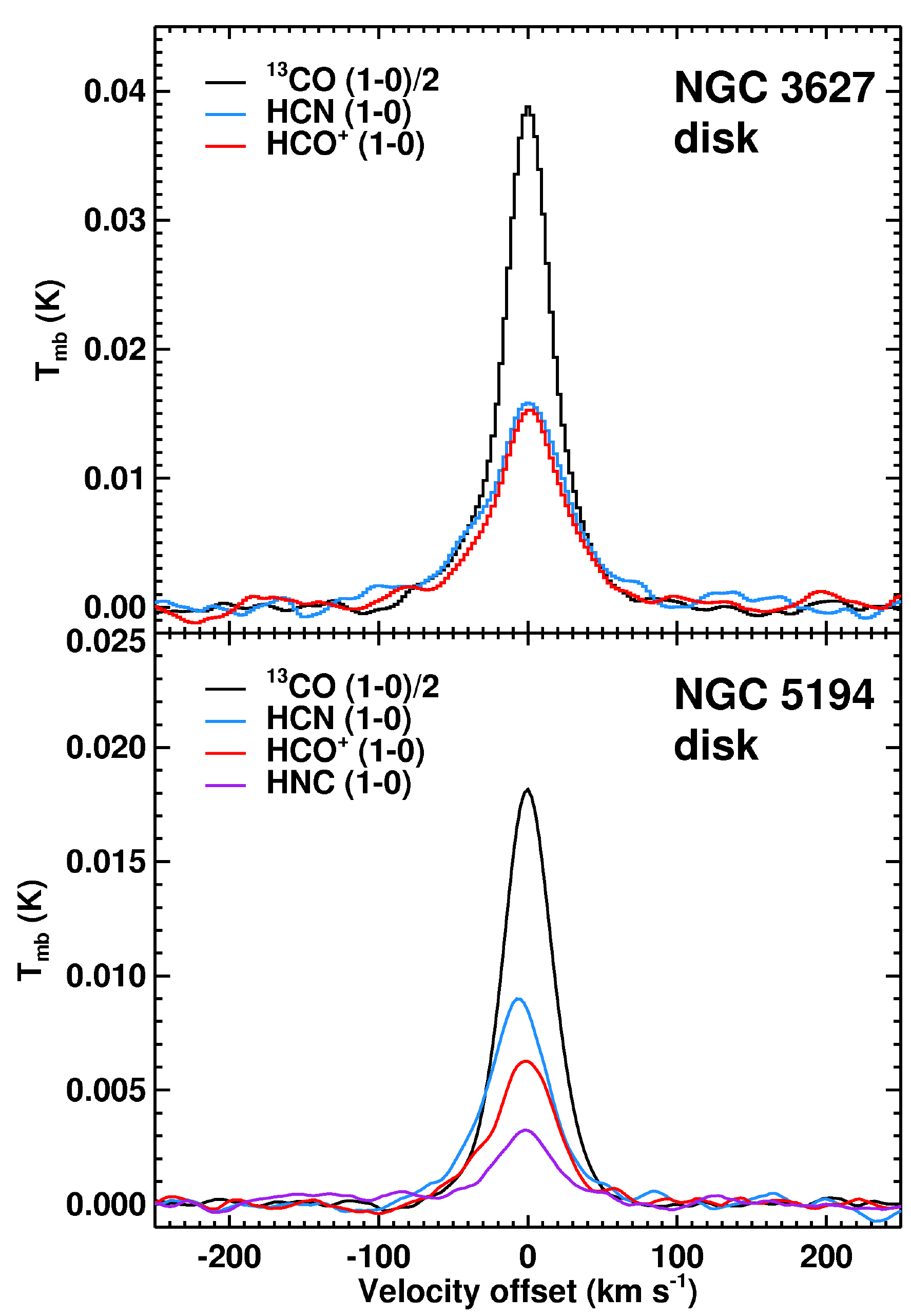}
\end{center}
\caption{Stacked spectra across the entire molecular disc for NGC~3627 and NGC~5194 excluding the central aperture (Figures \ref{densegas_maps} and \ref{centers}). We show spectra for the main dense gas tracers together with the stacked $^{13}$CO spectrum.} Line widths are larger by up to a factor of $\sim$2 in the inner parts compared to the discs, which we attribute largely to unresolved bulk motions.
\label{disks-width}
\end{figure}

In addition to line ratios, the stacked spectra allow us to compare line widths among the different molecular lines. Potential differences in the line widths may hint at differences in the distribution of such emission along the line of sight.

Table \ref{table:lines} and Figure \ref{centers} show that we do not observe strong differences among the line widths of the dense gas tracers and the $^{13}$CO emission tracing lower density gas. Thus, on the scale of our beam, which ranges from a few hundred pc to $\approx 1$~kpc (except for the much higher resolution NGC\,253), dense gas tracers appear well-mixed with $^{13}$CO. Note that at these scales, we expect each beam to encompass many individual clouds and that we stack beam-by-beam. Therefore the statement here supports the idea that the inter-cloud line width for HCN and HCO$^+$ resembles that for $^{13}$CO. There does not appear to be a large population of clouds emitting only $^{13}$CO that show different velocities than the clouds emitting dense gas tracers.

We do observe spatial variations in the line width. When exploring stacked emission from discs, which we mostly omit from this paper, we find line widths as much as a factor of $\approx 2$ narrower than in galaxy centres (see Figures \ref{centers} and \ref{disks-width}). And in the high spatial resolution data that we use to explore NGC\,253, we find narrower line widths than in our other targets despite the high inclination and high degree of turbulence in this galaxy. In NGC\,5194 the difference between the disc line width ($\sim 48$~km~s$^{-1}$) and the central line width ($\sim 70$~km~s$^{-1}$) is almost a factor of two and is present for all lines. The simplest explanation for this difference, and a similar difference shown for NGC\,3627, is that the broad line widths in the central region of the galaxy still contain large amounts of unresolved bulk motion. We expect this to mostly be rotation unresolved by our coarse beam (i.e., ``beam smearing'') but this may also include streaming motions along the spiral arms and bar or even some contributions from the AGN \citep[e.g., see ][]{1998ApJ...493L..63S,2013ApJ...779...45M,2015PKAS...30..439M,2016arXiv160700010Q}. The narrower line widths in NGC~253 likely reflect the fact that we include less large scale dynamical motion and perhaps that we sample the turbulent cascade at an intermediate scale \citep[e.g.][]{2016AJ....151...34C}.

\section{Discussion}
\label{discussion}

Figure \ref{comparison} shows our constraints on isotopologue line ratios for HCN and HCO$^+$. The $^{12}$C-to-$^{13}$C line ratios tend to be higher than those observed for CO molecules, but lower than typical isotopic ratios. How should we interpret these? A main motivation for this study was to constrain the optical depth of the dense gas tracers, and so to understand the importance of line trapping. In the simple case of a fixed $^{12}$C/$^{13}$C abundance and a cloud with a single density and uniform abundances, these ratios can be translated to an optical depth. The ratios that we observe tend to be lower than commonly inferred $^{12}$C/$^{13}$C abundances, implying some optical depth in the $^{12}$C lines. This section steps through key assumptions and results for this simple model. We then discuss how a breakdown in these simple assumptions may drive our observed line ratios. In particular, in the case where the $^{12}$C lines are optically thick then we would expect the $^{12}$C and $^{13}$C lines to emerge from distinct regions of the clouds.

\subsection{Carbon Isotope Abundance Ratio}
\label{sec:carbon}

Table \ref{table:carbon} lists estimate of the $^{12}$C/$^{13}$C abundance ratio. A ratio of 89 is thought to characterise the Solar System at its formation \citep{1994ARA&A..32..191W}, while local clouds show $\sim 40$--$60$, the Milky Way centre shows a low ratio of $\sim 25$, and starburst galaxies at low and high redshift can reach ratios of $\sim 100$. This value varies, at least in part, because $^{12}$C and $^{13}$C are produced by different phases of stellar evolution. $^{13}$C is only created primarily as a result of evolved intermediate-mass stars, whereas very massive stars are the ones that produce $^{12}$C at the end of their lives. As a consequence, the $^{12}$C/$^{13}$C ratio depends on the recent star formation history, tending to be high in regions of recent star formation and then dropping with time.

We measure ratios for the central regions of spiral galaxies. These tend to be actively star-forming regions, but not as extreme as those in ULIRGs. Based on the range of literature values, we adopt $50$ as our fiducial $^{12}$C/$^{13}$C ratio, since its determination is not possible with the present data set, but we note that this is likely uncertain by a factor of $\sim 2$. We consider this effect in our EMPIRE and ALMA sample in Jim\'enez-Donaire et al. (in prep.). Several effects can affect this ratio, like selective photo-dissociation or isotopic fractionation. While the strong interstellar radiation field emitted from OB stars, especially in strong starbursts, does not dissociate $^{12}$CO in large amounts due to self-shielding \citep[e.g.][]{1988ApJ...334..771V}, $^{13}$CO may be more severely affected \citep{1982ApJ...255..143B} which may increase the $^{12}$C/$^{13}$C ratio. On the other hand, chemical fractionation in cold regions would lead to preferential formation of $^{13}$CO.
Qualitatively, fractionation effects would likely marginally increase the H$^{13}$CO$^+$ abundance \citep{1990ApJ...357..477L, 2005ApJ...634.1126M}, whereas H$^{13}$CN and HN$^{13}$C are expected to decrease \citep[e.g.][]{2015A&A...576A..99R}.

\begin{table}
\caption{Carbon isotope ratios from the literature in different environments}           
\label{table:carbon}      
\centering                          
\begin{tabular}{l c r c}        

\hline\hline                 
 Type of Object & Galaxy & $^{12}$C/$^{13}$C & Ref.\\
\hline                        

Centre & Milky Way & $\sim$25 & 1\\
LMC & Magellanic Clouds & $\sim$50 & 2\\
NGC 2024 & Milky Way & $\sim$65 & 3\\
Orion A & Milky Way & $\sim$45 & 3\\
Starburst & NGC 253 & $\sim$40 & 4\\
  &  & $\sim$80 & 5\\
Starburst & M 82 & $>$ 40 & 4\\
  &  & $>$ 130 & 5\\
Local ULIRGs & Mrk 231/Arp 220 & $\sim$100 & 4, 6\\
ULIRG, $z=2.5$ & Cloverleaf & $>$ 100 & 7\\
\hline                                   
\end{tabular}
\\(1) \citet{1985A&A...149..195G}; (2) \citet{2009ApJ...690..580W}; (3) \citet{2002ApJ...578..211S}; (4) \citet{2014A&A...565A...3H}; (5) \citet{2010A&A...522A..62M}; (6) \citet{2012A&A...541A...4G}; (7) \citet{2010A&A...516A.111H}.
\end{table}

\subsection{Optical Depth in the Simple Case}
\label{sec:opacities}

\subsubsection{Framework}

In the simple case of a cloud with a single density, $^{12}$C and $^{13}$C lines evenly mixed throughout the cloud, and the $^{12}$C line optically thick, then the line ratio implies an optical depth. 

To estimate this optical depth, we begin in the following with the simplest case of assuming local thermodynamic equilibrium (LTE) and in particular $T_{\textnormal{ex}}$(HCN)$=T_{\textnormal{ex}}$(H$^{13}$CN). Note that we will drop this assumption in Section \ref{sec:changes}. We consider the intensity, $J_\nu$, of the $^{12}$C and $^{13}$C lines in units of brightness temperature, but not necessarily on the Rayleigh-Jeans tail, so that:

\begin{equation}
\label{eq:tb}
J_\nu=\frac{h\nu/k}{e^{h\nu/kT}-1}.
\end{equation}

\noindent where $\nu$ is the observed frequency, $k$ is Boltzmann's constant, $h$ is Planck's constant, and $T$ is the temperature characterising the source function. For our cloud, $T$ will be $T_{\textnormal{ex}}$ in LTE. Then the equation of radiative transfer gives the observed intensity, $T_{\rm obs}$, in units of brightness temperature:

\begin{equation}
\label{rad}
T_{\rm obs} = \eta_{\textnormal{bf}} \left[J_\nu(T_\textnormal{ex})-J_\nu(T_\textnormal{bg})\right] \left(1-\exp{(-\tau)}\right)~.
\end{equation}

\noindent Here $\eta_{\textnormal{bf}}$ is the beam filling factor, the fraction of the beam area that is filled by the emitting region. $T_\textnormal{bg}$ is the background radiation field temperature ($2.71$~K), and $\tau$ is the optical depth of the transition in question. Note that $\eta_{\rm bf}$, the beam filling factor, is expected to be a very small number, bringing our observed intensities from $\sim 10$~K down to the observed small fraction of a Kelvin.

If we make the simplifying assumption that the $^{12}$C lines, e.g., HCN~(1-0), are optically thick, then $1 - \exp{(-\tau)} \approx 1$ and 

\begin{equation}
\label{beam}
T_{\rm obs}^{12C} = \eta_{\textnormal{bf}}^{12C} \left[J_{\nu,12C}(T_\textnormal{ex})-J_{\nu,12C}(T_\textnormal{bg})\right],
\end{equation}

\noindent where $J_{\nu,12C}$ refers to Equation \ref{eq:tb} evaluated at the frequency of the $^{12}$C line. On the other hand, the $^{13}$C line has unknown optical depth, so that:

\begin{equation}
T_{\rm obs}^{13C} = \eta_{\textnormal{bf}}^{13C} \left[J_{\nu,13C}(T_\textnormal{ex})-J_{\nu,13C}(T_\textnormal{bg})\right]\left(1-\exp{(-\tau^{13C})}\right)~.
\end{equation}

\noindent Here we have distinguished the frequency of the $^{13}$C line from that of the $^{12}$C line above. The equations also allow that the two lines might have different filling factors and so distinguish $\eta_{\rm bf}^{13C}$ from $\eta_{\rm bf}^{12C}$. The optical depth of the $^{13}$C line is unknown. However, our observed ratios are far from $\sim 1$, which would be expected for both lines optically thick and matched $\eta_{\rm bf}$. Thus, it appears unlikely that both lines are optically thick. Therefore we expect $\left(1-\exp{-\tau^{13C}}\right) \approx \tau^{13C}$ in most cases.

Then, generally for LTE and an optically thick $^{12}$C line we have:

\begin{equation}
\label{temperatures}
\frac{T_{\rm obs}^{13C}}{T_{\rm obs}^{12C}}=\frac{\eta_{\textnormal{bf}}^{13C}}{\eta_{\textnormal{bf}}^{12C}}\,\frac{J_{\nu,13C}(T_\textnormal{ex})-J_{\nu,13C}(T_\textnormal{bg})}{J_{\nu,12C}(T_\textnormal{ex})-J_{\nu,12C}(T_\textnormal{bg})} \left(1-e^{-\tau^{13C}}\right)~.
\end{equation}

The differences between the $^{12}$C and $^{13}$C frequencies are small, so that

\begin{equation}
\label{temperatures}
\frac{T_{\rm obs}^{13C}}{T_{\rm obs}^{12C}} \approx  \frac{\eta_{\textnormal{bf}}^{13C}}{\eta_{\textnormal{bf}}^{12C}} \left( 1-e^{-\tau^{13C}} \right)
\end{equation}

In the simple case, we further assume that the $^{13}$C line is optically thin and that the beam filling factors of the two lines match (an assumption we will drop in the next section), so that:

\begin{equation}
\label{tau13_final}
\tau^{13C}  = -\ln\left(1-\frac{T_{\rm obs}^{13C}}{T_{\rm obs}^{12C}}\right) \approx \frac{T_{\rm obs}^{13C}}{T_{\rm obs}^{12C}}
\end{equation}

This exercise yields the optical depth of the $^{13}$C line. Because the opacity is proportional to the column of molecules, we can relate $\tau^{13C}$ to $\tau^{12C}$. Again ignoring minor differences between the lines, the ratio of optical depths will simply be the ratio in abundances between the two molecules,

\begin{equation}
\label{tau12}
\tau^{12C}=\tau^{13C} \,\frac{[\textrm{H}^{12}\textrm{CN}]}{[\textrm{H}^{13}\textrm{CN}]} \approx \tau^{13C} \frac{[^{12}\textrm{C}]}{[^{13}\textrm{C}]}~.
\end{equation}

\subsubsection{Results}

Using the line ratios tabulated in Table \ref{table:ratios}, Equations \ref{tau13_final} and \ref{tau12}, and adopting a Carbon isotope ratio of $^{12}$C/$^{13}$C$=50$, we estimate $\tau^{13C}$ and $\tau^{12C}$ for each pair of isotopologues. We report these in Table \ref{table:depths}. Because they use Equation \ref{tau13_final}, these estimates all assume matched beam filling factors and neglect any differences between the frequencies of the two transitions.

In Section \ref{sec:method}, we place upper limits on the $^{13}$C/$^{12}$C ratio in a number of cases. These translate into upper limits on $\tau^{12C}$ and $\tau^{13C}$. Broadly, the $^{13}$C lines are consistent with being optically thin. In this simple case, all of these optical depths are $\lesssim 0.3$. The corresponding optical depths for the $^{12}$C lines are all consistent with $\tau^{12C} \gtrsim 1$, in agreement with our assumption of thick $^{12}$C emission. In the case of HCN emission for NGC~3351, NGC~3627, and NGC~253 and HCO$^+$ for NGC~253, these are all consistent with $\tau^{12C} \sim 3$ \citep[see also][]{2015ApJ...801...63M}. However, in the other cases, these results are upper limits. For $^{13}$C upper limits, we can mainly conclude that the ratios do not rule out optically thick $^{12}$C emission.

Thus, in the simple case, the line ratios strongly suggest thin $^{13}$C emission. They are consistent with moderate optical depth in the $^{12}$C lines.

\begin{table*}
\caption{Optical depths and derived effective critical densities, assuming matching beam filling factors and a common excitation temperature for the $^{12}$C and $^{13}$C isotopologues. Critical densities are in units of cm$^{-3}$.}
\label{table:depths}
\centering
\begin{tabular}{l r r r r r r}        
\hline\hline                 
Galaxy &NGC 3351& NGC 3627& NGC 4254 & NGC 4321 &NGC 5194 & NGC 253\\
\hline                        

$\tau(\textrm{H}^{13}\textrm{CN})$ & 0.04$\pm$0.02 & 0.15$\pm$0.07 & $<$ 0.2 & $<$ 0.05 & $<$ 0.07 & 0.07$\pm$0.01\\

$\tau(\textrm{HCN})$ & 2.4$\pm$0.4 & 4.2$\pm$0.5 & $<$ 11 & $<$ 2.3& $<$ 3.5 & 2.5$\pm$0.6 \\

$\tau(\textrm{H}^{13}\textrm{CO}^{+})$ & $<$ 0.05 & $<$ 0.1 & $<$ 0.1 & $<$ 0.06 & $<$ 0.1 & 0.05$\pm$0.02\\

$\tau(\textrm{HCO}^{+})$ & $<$ 2.5 & $<$ 5.2 & $<$ 6.8 & $<$ 2.8 & $<$ 6.1 & 1.8$\pm$0.9 \\

$\tau(\textrm{HN}^{13}\textrm{C})$ & -- & -- & -- & -- & $<$ 0.3 & --\\

$\tau(\textrm{HNC})$ & -- & -- & -- & -- & $<$ 11 & -- \\

$\text{n}_\text{thick}(\textrm{HCN})$ & (2.0$\pm$0.2)$\times10^6$ & $>$ 8.6$\times10^5$ & $>$ 4.8$\times10^5$ & $>$ 2.0$\times10^6$ & $>$ 1.4$\times10^6$ & (2.0$\pm$0.2)$\times10^6$ \\
$\text{n}_\text{thick}(\textrm{HCO$^+$})$ & $>$ 2.9$\times10^5$ & $>$ 1.5$\times10^5$ & $>$ 1.1$\times10^5$ & $>$ 2.5$\times10^5$ & $>$ 1.2$\times10^5$ & (4.2$\pm$0.3)$\times10^5$ \\
$\text{n}_\text{thick}(\textrm{HNC})$ & -- & -- & -- & -- & $>$ 8.6$\times10^4$ & -- \\
\hline                  
\end{tabular}
\end{table*}

\subsection{Non-LTE Effects and Variable Beam Filling Factors}
\label{sec:changes}

In the previous section, we assumed that $\eta_{\rm bf}^{12C} = \eta_{\rm bf}^{13C}$, so that both lines emerge from the same region. In fact, the $^{12}$C line can be optically thick while the $^{13}$C line is optically thin. In this case, line trapping effects will lower the effective density for emission for the $^{12}$C line but not the $^{13}$C line. Then, lower density gas can emit strongly in the $^{12}$C but more weakly in the $^{13}$C line. This should yield $\eta_{\rm bf}^{12C} > \eta_{\rm bf}^{13C}$. That is, the two lines will show distinct distributions within a cloud, with the $^{13}$C line (which has no line trapping) confined to the denser parts of the cloud.

The same effect may lead to different excitation temperatures for the two lines. If the $^{12}$C line is optically thick, then line trapping will lower its effective critical density. If the collider (H$_2$) volume density in the galaxy is high compared to the critical density of the $^{12}$C line, but low compared to the critical density of the $^{13}$C line, then the $^{13}$C line may be sub-thermally excited while the $^{12}$C line approaches LTE. In this case, $T_{\rm ex}^{13C} < T_{\rm ex}^{12C}$.

The strength of non-LTE effects depends on the exact density distribution within the emitting region (see A. Leroy et al., ApJ submitted). If the region has mostly gas above the critical density of both lines, then the LTE case discussed above will apply. This might often be the case, e.g., for $^{12}$CO and $^{13}$CO.

On the other hand, if the density of the region is small compared to the effective critical density of the $^{12}$C line, then neither line will emit effectively. Most collisional excitations will be balanced by radiative de-excitations and the line ratio will drop. In the extreme limit, the line ratio will reach the abundance ratio.

In this paper, we observe mostly the central, bright areas of galaxies, focusing on regions with bright HCN emission. We do not expect these central zones to be dominated by extremely low volume density gas. However, if a large amount of $n_{\rm H2} \sim 10^{3.5}{-}10^4$~cm$^{-3}$ gas contributes to the emission, then these non-LTE effects may still be important. This is the range of densities where $n_{\rm H2}$ might be above the effective critical density for an optically thick $^{12}$C line but not an optically thin $^{13}$C line. This range of densities is also well within the range commonly observed inside molecular clouds in nuclear star forming regions. 

A. Leroy et al. (submitted) explore these effects for realistic density distributions. They calculate the amount by which the $^{12}$C to $^{13}$C line ratio will be suppressed for different lines given some optical depth of the $^{12}$C line and an adopted density distribution. The magnitude of the effect depends on the line, the optical depth, and the density distribution. For a log-normal distribution with a power-law tail, they found that the $^{12}$C/$^{13}$C line ratio could be enhanced relative to the LTE expectation by a factor of $\sim 2{-}3$ for HCN or HCO$^+$ over a range of plausible densities.

This enhanced line ratio can be captured to first order, by applying different filling factors to the two lines. Thus, to explore the implication of these effects, we revisit our LTE calculations but now assume that $\eta_{\rm bf}^{12C} / \eta_{\rm bf}^{13C} \approx 2$. To do this, we return to Equation \ref{temperatures}. We no longer set the beam filling factors equal between the two lines, but still set the temperatures and frequencies equal, and assume LTE with a single common temperature. Then:

\begin{equation}
\label{ffactor_eq}
\tau^{13C}_{\rm obs} \backsimeq -\ln\left(1-\frac{\eta_{\rm bf}^{12C}}{\eta_{\rm bf}^{13C}}\frac{T_{\rm obs}^{13C}}{T_{\rm obs}^{12C}}\right)~.
\end{equation}

\noindent This is still an approximation, but one that captures some of the effect of differential excitation of the $^{12}$C and $^{13}$C lines in the non-LTE case. We refer the reader to the non-LTE treatment in Leroy et al. (submitted) for more details and a tabulation of a variety of cases.

Equation \ref{ffactor_eq} shows that for $\eta_{\rm bf}^{12C} > \eta_{\rm bf}^{13C}$, our simple estimates of $\tau$ in the previous section will be underestimates. We report revised values in Table \ref{table:changes_tau}. These are higher than the corresponding LTE values in Table \ref{table:depths}. 

Even accounting for differential filling factors, we still derive $\tau^{13C}$ consistent with optically thin $^{13}$C emission. These non-LTE effects tend to make the limits on optical depth less stringent because they can push the line ratio towards ``optically thin'' values even while the $^{12}$C line remains thick. Accounting for this effect, our observations may still accommodate substantial thickness in the $^{12}$C line, with $\tau^{12C}$ potentially as high as $\sim 10$ for our HCN detections. 

The exact magnitude of any non-LTE correction will depend on the sub-beam density distribution, the true temperature of the gas, and the optical depth of the lines in question. Our best guess is that a moderate correction does indeed apply, and the factor of $\sim 2$ correction applied in Table \ref{table:changes_tau} may represent a reasonable estimate. In fact, if most of the emission that we see comes from dense clouds, this may be an over-correction, while if no power law tail is present, the correction for non-LTE effects may be even higher. Future multi-line work and isotopologue studies of Galactic Centre clouds will help refine these estimates further.

\begin{table*}
\caption{Optical depths and effective critical densities when considering $\eta_\text{12}\sim 2\, \eta_\text{13}$ and under the assumption of a common $T_\text{ex}$. Critical densities are in units of cm$^{-3}$.}

\label{table:changes_tau}      
\centering                          
\begin{tabular}{l c c c c c c}        

\hline\hline                 
 Target & $\tau_\textnormal{HCN}$ & n$_\text{crit}^{\text{HCN}}$ & $\tau_{\textnormal{HCO}^+}$ & n$_\text{crit}^{\text{HCO}^+}$ & $\tau_\textnormal{HNC}$ & n$_\text{crit}^{\text{HNC}}$\\

\hline                        

NGC 3351 & 5.0$\pm$0.4 & (1.0$\pm$0.4)$\times 10^{6}$ & $<$5.3 & $>$1.4$\times 10^{5}$ & -- & --\\
NGC 3627 & 8.8$\pm$0.7 & (5.6$\pm$0.1)$\times 10^{5}$ & $<$10.6 & $>$7.0$\times 10^{4}$ & -- & --\\
NGC 4254 & $<$25.8 & $>$1.2$\times 10^{5}$ & $<$16.0 & $>$4.5$\times 10^{4}$ & -- & --\\
NGC 4321 & $<$5.3 & $>$9.4$\times 10^{5}$ & $<$5.8 & $>$1.3$\times 10^{5}$ & -- & --\\
NGC 5194 & $<$8.0 & $>$6.2$\times 10^{5}$ & $<$14.3 & $>$5.1$\times 10^{4}$ & $<$31 & $>$1.8$\times 10^{4}$ \\
NGC 253 & 6.3$\pm$0.5 & (7.9$\pm$0.1)$\times 10^{5}$ & 4.3$\pm$0.6 & (1.7$\pm$0.1)$\times 10^{5}$ & -- & --\\

\hline                                   
\end{tabular}
\end{table*}

\subsection{Implications for Dense Gas Tracers}

Our simple calculations imply moderate optical depth for the dense gas tracers (the $^{12}$C lines) and this depth increases in the case of differential beam filling. Optical thickness in these lines means that radiative trapping will lead to emitted photons being re-absorbed. This lowers the effective critical density and changes the density at which these lines emit most effectively. This correction is likely to be significant for commonly used dense gas tracers. It will affect both the density required for effective emission and the conversion between line brightness and dense gas mass, the dense gas conversion factor.
 
In a review of this topic, \cite{10.1086/680342} note the effective critical density in the presence of line trapping as:

\begin{equation}
\label{new_crit}
n_\textnormal{crit}^{\textnormal{thick}} = \frac{\bar{\beta} \, A_{jk}}{\sum_{i\neq j}\gamma_{ji}}~.
\end{equation}

\noindent Here $\bar{\beta}$ is the solid angle-averaged escape fraction, $A_{jk}$ are the Einstein $A$ coefficients, and $\gamma_{ji}$ are the collision rates (cm$^3$ s$^{-1}$) out of level $j$ into another level $i$. The difference with the normal critical density calculation is that $\bar{\beta} \neq 1$, so that not every spontaneous emission escapes the cloud. In the simple case of spherical geometry and large optical depth ($\tau > 1$), the effective spontaneous transition rate becomes $\bar{\beta} \, A_{jk} \sim A_{jk}/\tau$, so that the critical density is depressed relative to its nominal (optically thin) value by a factor equal to the optical depth.

Tables \ref{table:depths} and \ref{table:changes_tau} report the effect of applying corrections based on the estimated optical depths to each line. The effect is to reduce the critical density by a factor of $\sim$2-6, yielding values that are typically $\sim 7\times10^5$ cm$^{-3}$ for HCN, $\sim 1\times10^5$ cm$^{-3}$ for HCO$^+$ and $> 1.8\times10^4$ cm$^{-3}$ for HNC. These are, of course, as uncertain as the optical depths, but give some guide as to the density of gas traced by these lines. 

\section{Summary and conclusions}
\label{summary}

We present observations of the 1-0 transitions of the dense, molecular gas tracers HCN, HCO$^+$, HNC and their isotopologues H$^{13}$CN, H$^{13}$CO$^+$ and HC$^{13}$N across the discs of six nearby galaxies. These include IRAM 30m observations of NGC 5194 (M51), as part of the IRAM large program EMPIRE, ALMA observations of NGC 3351, NGC 3627, NGC 4254, NGC 4321 and NGC 253. Given the faint nature of the $^{13}$C isotopologues, we focus our analysis on the inner ($\sim$50" for M51 and $\sim$10" for the ALMA galaxies), high surface density regions of these galaxies and stack all spectra in this region for each galaxy to improve the significance of our measurements. We use this data set to constrain the optical depth of these lines and study implications for the effective critical densities. 

\begin{itemize}

\item We detect HCN, HCO$^+$, HNC (1-0) with high significance for each galaxy ($>5\sigma$) in the stacked spectra. Emission from these lines is in fact well detected for individual lines of sight and resolved across the discs of our sample, and in particular, the HCO$^+$ and HNC emission are distributed very similarly to the HCN emission in NGC\,5194 \citep{2016ApJ...822L..26B}. The distribution of these dense gas tracers follows well the bulk molecular medium one as traced by CO emission; it is brightest in the inner parts and follows the spiral structure (where present) in CO.
\medskip
\item The H$^{13}$CN, H$^{13}$CO$^+$ and HN$^{13}$C (1-0) lines, however, are much fainter; in the stacked spectra from the inner regions of our sample, we detect H$^{13}$CN (1-0) in NGC\,3351, NGC\,3627 and NGC\,253, and H$^{13}$CO$^+$ (1-0) in NGC\,253. NGC\,4254, NGC\,4321 and NGC\,5194 show no isotopologue detection in their inner part. Therefore we derive stringent upper limits for those cases, which we use for our analysis.
\medskip
\item We use the optically thin isotopologues to compute or constrain optical depths for each molecular lines in the inner parts of our target galaxies and, where detected, also in selected CO and HCN bright disc positions. We find that HCN and HCO$^+$ have optical depths in the range $\sim$2-11 in the inner parts of the galaxies analysed (assuming a value of 50 for the $^{12}$C/$^{13}$C abundance ratio). HN$^{13}$C data is only available for NGC\,5194, where we derived an upper limit to the opacity for HCN in its inner region, $\tau < 11$. The optical depth for H$^{13}$CN, H$^{13}$CO$^+$ and HN$^{13}$C shows a larger degree of variation, ranging from 0.04 to 0.3 in the inner part of NGC\,3351, NGC\,3627 and NGC\,253. We conclude that HCN, HCO$^+$ and HNC (1-0) emission is largely moderately optically thick with optical depths of typically a few, whereas the H$^{13}$CN, H$^{13}$CO$^+$ and HN$^{13}$C (1-0) lines are largely optically thin, even in the inner parts of our galaxies on the $\sim$0.5-2 kpc scales we probe.
\medskip
\item Given their non-negligible optical thickness, the critical densities of the HCN, HCO$^+$ and HNC are reduced by a factor of 2-6, which implies that these lines are sensitive to molecular gas at lower densities. We also study the non-LTE conditions and the influence of variable beam filling factors for the different isotopologues: given their optical thickness, and thus sensitivity to lower density gas, one may expect larger beam filling factors for the $^{12}$C molecules. This would further increase optical depths and therefore lead to a further decrease in the effective critical densities for the optically thick dense gas tracers.
\medskip
\item We compare the HCN/H$^{13}$CN, HCO$^+$/H$^{13}$CO$^+$ and HNC/HN$^{13}$C line ratios measured in our sample to those compiled from the literature. There is good agreement between the values we find for NGC\,5194 and NGC\,253 with previous studies. The work done in the Galactic Centre also shows compatible values with those we find in the inner parts of our sample, within the uncertainties.

\end{itemize}

\section*{Acknowledgements}

The authors thank Daniel Harsono, Ralf Klessen, Fabian Walter, Simon Glover and Kazimierz Sliwa for discussions that have improved this manuscript. We thank the referee for detailed and helpful comments. FB, MJJD and DC acknowledge support from DFG grant BI 1546/1-1. AKL thanks the NRAO/University of Virginia star formation group for helpful discussions. The National Radio Astronomy Observatory is a facility of the National Science Foundation operated under cooperative agreement by Associated Universities, Inc. DC is supported by the Deutsche Forschungsgemeinschaft, DFG, through project number SFB956C. SGB acknowledges support from Spanish grants ESP2015-68964-P and AYA2013-42227-P. AH acknowledges support from the Centre National d'Etudes Spatiales (CNES). MRK acknowledges support from ARC grant DP160100695. ER is supported by a Discovery Grant from NSERC of Canada. AU acknowledges support from Spanish MINECO grants FIS2012-32096 and ESP2015-68964.


\bibliographystyle{mn2e}
\setlength{\bibhang}{2.0em}
\setlength\labelwidth{0.0em}
\bibliography{mj_bib}

\begin{thebibliography}{62}
\expandafter\ifx\csname natexlab\endcsname\relax\def\natexlab#1{#1}\fi

\bibitem[{{Aladro} {et~al}\mbox{.}(2013){Aladro}, {Viti}, {Bayet}, {Riquelme},
  {Mart{\'{\i}}n}, {Mauersberger}, {Mart{\'{\i}}n-Pintado}, {Requena-Torres},
  {Kramer}, \& {Wei{\ss}}}]{2013A&A...549A..39A}
{Aladro} R. {et~al.}, 2013, \aap, 549, A39

\bibitem[{{Andr{\'e}} {et~al}\mbox{.}(2014){Andr{\'e}}, {Di Francesco},
  {Ward-Thompson}, {Inutsuka}, {Pudritz}, \& {Pineda}}]{2014prpl.conf...27A}
{Andr{\'e}} P., {Di Francesco} J., {Ward-Thompson} D., {Inutsuka} S.-I.,
  {Pudritz} R.~E., {Pineda} J.~E., 2014, Protostars and Planets VI, 27

\bibitem[{{Bally} \& {Langer}(1982)}]{1982ApJ...255..143B}
{Bally} J., {Langer} W.~D., 1982, \apj, 255, 143

\bibitem[{{Battisti} \& {Heyer}(2013)}]{2013AAS...22134909B}
{Battisti} A., {Heyer} M.~H., 2013, in American Astronomical Society Meeting
  Abstracts, Vol. 221, American Astronomical Society Meeting Abstracts \#221,
  p. 349.09

\bibitem[{{Bigiel} {et~al}\mbox{.}(2016){Bigiel}, {Leroy},
  {Jim{\'e}nez-Donaire}, {Pety}, {Usero}, {Cormier}, {Bolatto},
  {Garcia-Burillo}, {Colombo}, {Gonz{\'a}lez-Garc{\'{\i}}a}, {Hughes},
  {Kepley}, {Kramer}, {Sandstrom}, {Schinnerer}, {Schruba}, {Schuster},
  {Tomicic}, \& {Zschaechner}}]{2016ApJ...822L..26B}
{Bigiel} F. {et~al.}, 2016, \apjl, 822, L26

\bibitem[{{Bolatto} {et~al}\mbox{.}(2013){Bolatto}, {Warren}, {Leroy},
  {Walter}, {Veilleux}, {Ostriker}, {Ott}, {Zwaan}, {Fisher}, {Weiss},
  {Rosolowsky}, \& {Hodge}}]{2013Natur.499..450B}
{Bolatto} A.~D. {et~al.}, 2013, \nat, 499, 450

\bibitem[{{Buchbender} {et~al}\mbox{.}(2013){Buchbender}, {Kramer},
  {Gonzalez-Garcia}, {Israel}, {Garc{\'{\i}}a-Burillo}, {van der Werf},
  {Braine}, {Rosolowsky}, {Mookerjea}, {Aalto}, {Boquien}, {Gratier}, {Henkel},
  {Quintana-Lacaci}, {Verley}, \& {van der Tak}}]{2013A&A...549A..17B}
{Buchbender} C. {et~al.}, 2013, \aap, 549, A17

\bibitem[{{Bussmann} {et~al}\mbox{.}(2008){Bussmann}, {Narayanan}, {Shirley},
  {Juneau}, {Wu}, {Solomon}, {Vanden Bout}, {Moustakas}, \&
  {Walker}}]{2008ApJ...681L..73B}
{Bussmann} R.~S. {et~al.}, 2008, \apjl, 681, L73

\bibitem[{{Cald{\'u}-Primo} \& {Schruba}(2016)}]{2016AJ....151...34C}
{Cald{\'u}-Primo} A., {Schruba} A., 2016, \aj, 151, 34

\bibitem[{{Cald{\'u}-Primo} {et~al}\mbox{.}(2013){Cald{\'u}-Primo}, {Schruba},
  {Walter}, {Leroy}, {Sandstrom}, {de Blok}, {Ianjamasimanana}, \&
  {Mogotsi}}]{2013AJ....146..150C}
{Cald{\'u}-Primo} A., {Schruba} A., {Walter} F., {Leroy} A., {Sandstrom} K.,
  {de Blok} W.~J.~G., {Ianjamasimanana} R., {Mogotsi} K.~M., 2013, \aj, 146,
  150

\bibitem[{{Carter} {et~al}\mbox{.}(2012){Carter}, {Lazareff}, {Maier}, {Chenu},
  {Fontana}, {Bortolotti}, {Boucher}, {Navarrini}, {Blanchet}, {Greve}, {John},
  {Kramer}, {Morel}, {Navarro}, {Pe{\~n}alver}, {Schuster}, \&
  {Thum}}]{2012A&A...538A..89C}
{Carter} M. {et~al.}, 2012, \aap, 538, A89

\bibitem[{{Chin} {et~al}\mbox{.}(1998){Chin}, {Henkel}, {Millar}, {Whiteoak},
  \& {Marx-Zimmer}}]{1998A&A...330..901C}
{Chin} Y.-N., {Henkel} C., {Millar} T.~J., {Whiteoak} J.~B., {Marx-Zimmer} M.,
  1998, \aap, 330, 901

\bibitem[{{Christopher} {et~al}\mbox{.}(2005){Christopher}, {Scoville},
  {Stolovy}, \& {Yun}}]{2005ApJ...622..346C}
{Christopher} M.~H., {Scoville} N.~Z., {Stolovy} S.~R., {Yun} M.~S., 2005,
  \apj, 622, 346

\bibitem[{{Ciardullo} {et~al}\mbox{.}(2002){Ciardullo}, {Feldmeier}, {Jacoby},
  {Kuzio de Naray}, {Laychak}, \& {Durrell}}]{2002ApJ...577...31C}
{Ciardullo} R., {Feldmeier} J.~J., {Jacoby} G.~H., {Kuzio de Naray} R.,
  {Laychak} M.~B., {Durrell} P.~R., 2002, \apj, 577, 31

\bibitem[{{Dumouchel}, {Faure} \& {Lique}(2010){Dumouchel}, {Faure}, \&
  {Lique}}]{2010MNRAS.406.2488D}
{Dumouchel} F., {Faure} A., {Lique} F., 2010, \mnras, 406, 2488

\bibitem[{{Flower}(1999)}]{1999MNRAS.305..651F}
{Flower} D.~R., 1999, \mnras, 305, 651

\bibitem[{{Gao} \& {Solomon}(2004)}]{2004ApJ...606..271G}
{Gao} Y., {Solomon} P.~M., 2004, \apj, 606, 271

\bibitem[{{Garc{\'{\i}}a-Burillo} {et~al}\mbox{.}(2012){Garc{\'{\i}}a-Burillo},
  {Usero}, {Alonso-Herrero}, {Graci{\'a}-Carpio}, {Pereira-Santaella},
  {Colina}, {Planesas}, \& {Arribas}}]{2012A&A...539A...8G}
{Garc{\'{\i}}a-Burillo} S., {Usero} A., {Alonso-Herrero} A.,
  {Graci{\'a}-Carpio} J., {Pereira-Santaella} M., {Colina} L., {Planesas} P.,
  {Arribas} S., 2012, \aap, 539, A8

\bibitem[{{Gonz{\'a}lez-Alfonso} {et~al}\mbox{.}(2012){Gonz{\'a}lez-Alfonso},
  {Fischer}, {Graci{\'a}-Carpio}, {Sturm}, {Hailey-Dunsheath}, {Lutz},
  {Poglitsch}, {Contursi}, {Feuchtgruber}, {Veilleux}, {Spoon}, {Verma},
  {Christopher}, {Davies}, {Sternberg}, {Genzel}, \&
  {Tacconi}}]{2012A&A...541A...4G}
{Gonz{\'a}lez-Alfonso} E. {et~al.}, 2012, \aap, 541, A4

\bibitem[{{Graci{\'a}-Carpio} {et~al}\mbox{.}(2006){Graci{\'a}-Carpio},
  {Garc{\'{\i}}a-Burillo}, {Planesas}, \& {Colina}}]{2006ApJ...640L.135G}
{Graci{\'a}-Carpio} J., {Garc{\'{\i}}a-Burillo} S., {Planesas} P., {Colina} L.,
  2006, \apjl, 640, L135

\bibitem[{{Graci{\'a}-Carpio} {et~al}\mbox{.}(2008){Graci{\'a}-Carpio},
  {Garc{\'{\i}}a-Burillo}, {Planesas}, {Fuente}, \&
  {Usero}}]{2008A&A...479..703G}
{Graci{\'a}-Carpio} J., {Garc{\'{\i}}a-Burillo} S., {Planesas} P., {Fuente} A.,
  {Usero} A., 2008, \aap, 479, 703

\bibitem[{{Guesten}, {Henkel} \& {Batrla}(1985){Guesten}, {Henkel}, \&
  {Batrla}}]{1985A&A...149..195G}
{Guesten} R., {Henkel} C., {Batrla} W., 1985, \aap, 149, 195

\bibitem[{{Harada} {et~al}\mbox{.}(2015){Harada}, {Riquelme}, {Viti},
  {Jim{\'e}nez-Serra}, {Requena-Torres}, {Menten}, {Mart{\'{\i}}n}, {Aladro},
  {Martin-Pintado}, \& {Hochg{\"u}rtel}}]{2015A&A...584A.102H}
{Harada} N. {et~al.}, 2015, \aap, 584, A102

\bibitem[{{Heiderman} {et~al}\mbox{.}(2010){Heiderman}, {Evans}, {Allen},
  {Huard}, \& {Heyer}}]{2010ApJ...723.1019H}
{Heiderman} A., {Evans}, II N.~J., {Allen} L.~E., {Huard} T., {Heyer} M., 2010,
  \apj, 723, 1019

\bibitem[{{Henkel} {et~al}\mbox{.}(2014){Henkel}, {Asiri}, {Ao}, {Aalto},
  {Danielson}, {Papadopoulos}, {Garc{\'{\i}}a-Burillo}, {Aladro},
  {Impellizzeri}, {Mauersberger}, {Mart{\'{\i}}n}, \&
  {Harada}}]{2014A&A...565A...3H}
{Henkel} C. {et~al.}, 2014, \aap, 565, A3

\bibitem[{{Henkel} {et~al}\mbox{.}(1998){Henkel}, {Chin}, {Mauersberger}, \&
  {Whiteoak}}]{1998A&A...329..443H}
{Henkel} C., {Chin} Y.-N., {Mauersberger} R., {Whiteoak} J.~B., 1998, \aap,
  329, 443

\bibitem[{{Henkel} {et~al}\mbox{.}(2010){Henkel}, {Downes}, {Wei{\ss}},
  {Riechers}, \& {Walter}}]{2010A&A...516A.111H}
{Henkel} C., {Downes} D., {Wei{\ss}} A., {Riechers} D., {Walter} F., 2010,
  \aap, 516, A111

\bibitem[{{Kennicutt} \& {Evans}(2012)}]{2012ARA&A..50..531K}
{Kennicutt} R.~C., {Evans} N.~J., 2012, \araa, 50, 531

\bibitem[{{Krips} {et~al}\mbox{.}(2008){Krips}, {Neri},
  {Garc{\'{\i}}a-Burillo}, {Mart{\'{\i}}n}, {Combes}, {Graci{\'a}-Carpio}, \&
  {Eckart}}]{2008ApJ...677..262K}
{Krips} M., {Neri} R., {Garc{\'{\i}}a-Burillo} S., {Mart{\'{\i}}n} S., {Combes}
  F., {Graci{\'a}-Carpio} J., {Eckart} A., 2008, \apj, 677, 262

\bibitem[{{Lada} \& {Lada}(2003)}]{2003ARA&A..41...57L}
{Lada} C.~J., {Lada} E.~A., 2003, \araa, 41, 57

\bibitem[{{Lada}, {Lombardi} \& {Alves}(2010){Lada}, {Lombardi}, \&
  {Alves}}]{2010ApJ...724..687L}
{Lada} C.~J., {Lombardi} M., {Alves} J.~F., 2010, \apj, 724, 687

\bibitem[{{Langer} \& {Penzias}(1990)}]{1990ApJ...357..477L}
{Langer} W.~D., {Penzias} A.~A., 1990, \apj, 357, 477

\bibitem[{{Leroy} {et~al}\mbox{.}(2015){Leroy}, {Bolatto}, {Ostriker},
  {Rosolowsky}, {Walter}, {Warren}, {Donovan Meyer}, {Hodge}, {Meier}, {Ott},
  {Sandstrom}, {Schruba}, {Veilleux}, \& {Zwaan}}]{2015ApJ...801...25L}
{Leroy} A.~K. {et~al.}, 2015, \apj, 801, 25

\bibitem[{{Leroy} {et~al}\mbox{.}(2013){Leroy}, {Walter}, {Sandstrom},
  {Schruba}, {Munoz-Mateos}, {Bigiel}, {Bolatto}, {Brinks}, {de Blok}, {Meidt},
  {Rix}, {Rosolowsky}, {Schinnerer}, {Schuster}, \&
  {Usero}}]{2013AJ....146...19L}
{Leroy} A.~K. {et~al.}, 2013, \aj, 146, 19

\bibitem[{{Marr}, {Wright} \& {Backer}(1993){Marr}, {Wright}, \&
  {Backer}}]{1993ApJ...411..667M}
{Marr} J.~M., {Wright} M.~C.~H., {Backer} D.~C., 1993, \apj, 411, 667

\bibitem[{{Mart{\'{\i}}n} {et~al}\mbox{.}(2010){Mart{\'{\i}}n}, {Aladro},
  {Mart{\'{\i}}n-Pintado}, \& {Mauersberger}}]{2010A&A...522A..62M}
{Mart{\'{\i}}n} S., {Aladro} R., {Mart{\'{\i}}n-Pintado} J., {Mauersberger} R.,
  2010, \aap, 522, A62

\bibitem[{{Matsushita} {et~al}\mbox{.}(2015){Matsushita}, {Trung}, {Boone},
  {Krips}, {Lim}, \& {Muller}}]{2015PKAS...30..439M}
{Matsushita} S., {Trung} D.-V., {Boone} F., {Krips} M., {Lim} J., {Muller} S.,
  2015, Publication of Korean Astronomical Society, 30, 439

\bibitem[{{McMullin} {et~al}\mbox{.}(2007){McMullin}, {Waters}, {Schiebel},
  {Young}, \& {Golap}}]{2007ASPC..376..127M}
{McMullin} J.~P., {Waters} B., {Schiebel} D., {Young} W., {Golap} K., 2007, in
  Astronomical Society of the Pacific Conference Series, Vol. 376, Astronomical
  Data Analysis Software and Systems XVI, {Shaw} R.~A., {Hill} F., {Bell}
  D.~J., eds., p. 127

\bibitem[{{Meidt} {et~al}\mbox{.}(2013){Meidt}, {Schinnerer},
  {Garc{\'{\i}}a-Burillo}, {Hughes}, {Colombo}, {Pety}, {Dobbs}, {Schuster},
  {Kramer}, {Leroy}, {Dumas}, \& {Thompson}}]{2013ApJ...779...45M}
{Meidt} S.~E. {et~al.}, 2013, \apj, 779, 45

\bibitem[{{Meier} {et~al}\mbox{.}(2015){Meier}, {Walter}, {Bolatto}, {Leroy},
  {Ott}, {Rosolowsky}, {Veilleux}, {Warren}, {Wei{\ss}}, {Zwaan}, \&
  {Zschaechner}}]{2015ApJ...801...63M}
{Meier} D.~S. {et~al.}, 2015, \apj, 801, 63

\bibitem[{{Milam} {et~al}\mbox{.}(2005){Milam}, {Savage}, {Brewster}, {Ziurys},
  \& {Wyckoff}}]{2005ApJ...634.1126M}
{Milam} S.~N., {Savage} C., {Brewster} M.~A., {Ziurys} L.~M., {Wyckoff} S.,
  2005, \apj, 634, 1126

\bibitem[{{Pety}(2005)}]{2005sf2a.conf..721P}
{Pety} J., 2005, in SF2A-2005: Semaine de l'Astrophysique Francaise, {Casoli}
  F., {Contini} T., {Hameury} J.~M., {Pagani} L., eds., p. 721

\bibitem[{{Pety} {et~al}\mbox{.}(2013){Pety}, {Schinnerer}, {Leroy}, {Hughes},
  {Meidt}, {Colombo}, {Dumas}, {Garc{\'{\i}}a-Burillo}, {Schuster}, {Kramer},
  {Dobbs}, \& {Thompson}}]{2013ApJ...779...43P}
{Pety} J. {et~al.}, 2013, \apj, 779, 43

\bibitem[{{Querejeta} {et~al}\mbox{.}(2016){Querejeta}, {Schinnerer},
  {Garc{\'{\i}}a-Burillo}, {Bigiel}, {Blanc}, {Colombo}, {Hughes}, {Kreckel},
  {Leroy}, {Meidt}, {Meier}, {Pety}, \& {Sliwa}}]{2016arXiv160700010Q}
{Querejeta} M. {et~al.}, 2016, ArXiv e-prints

\bibitem[{{Riquelme} {et~al}\mbox{.}(2010){Riquelme}, {Bronfman},
  {Mauersberger}, {May}, \& {Wilson}}]{2010A&A...523A..45R}
{Riquelme} D., {Bronfman} L., {Mauersberger} R., {May} J., {Wilson} T.~L.,
  2010, \aap, 523, A45

\bibitem[{{Roueff}, {Loison} \& {Hickson}(2015){Roueff}, {Loison}, \&
  {Hickson}}]{2015A&A...576A..99R}
{Roueff} E., {Loison} J.~C., {Hickson} K.~M., 2015, \aap, 576, A99

\bibitem[{{Savage} {et~al}\mbox{.}(2002){Savage}, {Apponi}, {Ziurys}, \&
  {Wyckoff}}]{2002ApJ...578..211S}
{Savage} C., {Apponi} A.~J., {Ziurys} L.~M., {Wyckoff} S., 2002, \apj, 578, 211

\bibitem[{{Schinnerer} {et~al}\mbox{.}(2013){Schinnerer}, {Meidt}, {Pety},
  {Hughes}, {Colombo}, {Garc{\'{\i}}a-Burillo}, {Schuster}, {Dumas}, {Dobbs},
  {Leroy}, {Kramer}, {Thompson}, \& {Regan}}]{2013ApJ...779...42S}
{Schinnerer} E. {et~al.}, 2013, \apj, 779, 42

\bibitem[{{Schirm} {et~al}\mbox{.}(2016){Schirm}, {Wilson}, {Madden}, \&
  {Clements}}]{2016ApJ...823...87S}
{Schirm} M.~R.~P., {Wilson} C.~D., {Madden} S.~C., {Clements} D.~L., 2016,
  \apj, 823, 87

\bibitem[{{Schruba} {et~al}\mbox{.}(2011){Schruba}, {Leroy}, {Walter},
  {Bigiel}, {Brinks}, {de Blok}, {Dumas}, {Kramer}, {Rosolowsky}, {Sandstrom},
  {Schuster}, {Usero}, {Weiss}, \& {Wiesemeyer}}]{2011AJ....142...37S}
{Schruba} A. {et~al.}, 2011, \aj, 142, 37

\bibitem[{{Scoville} {et~al}\mbox{.}(1998){Scoville}, {Yun}, {Armus}, \&
  {Ford}}]{1998ApJ...493L..63S}
{Scoville} N.~Z., {Yun} M.~S., {Armus} L., {Ford} H., 1998, \apjl, 493, L63

\bibitem[{Shirley(2015)}]{10.1086/680342}
Shirley Y.~L., 2015, Publications of the Astronomical Society of the Pacific,
  127, 299

\bibitem[{{Stephens} {et~al}\mbox{.}(2016){Stephens}, {Jackson}, {Whitaker},
  {Contreras}, {Guzm{\'a}n}, {Sanhueza}, {Foster}, \&
  {Rathborne}}]{2016ApJ...824...29S}
{Stephens} I.~W., {Jackson} J.~M., {Whitaker} J.~S., {Contreras} Y.,
  {Guzm{\'a}n} A.~E., {Sanhueza} P., {Foster} J.~B., {Rathborne} J.~M., 2016,
  \apj, 824, 29

\bibitem[{{Usero} {et~al}\mbox{.}(2015){Usero}, {Leroy}, {Walter}, {Schruba},
  {Garc{\'{\i}}a-Burillo}, {Sandstrom}, {Bigiel}, {Brinks}, {Kramer},
  {Rosolowsky}, {Schuster}, \& {de Blok}}]{2015AJ....150..115U}
{Usero} A. {et~al.}, 2015, \aj, 150, 115

\bibitem[{{van der Tak} {et~al}\mbox{.}(2007){van der Tak}, {Black},
  {Sch{\"o}ier}, {Jansen}, \& {van Dishoeck}}]{2007A&A...468..627V}
{van der Tak} F.~F.~S., {Black} J.~H., {Sch{\"o}ier} F.~L., {Jansen} D.~J.,
  {van Dishoeck} E.~F., 2007, \aap, 468, 627

\bibitem[{{van Dishoeck} \& {Black}(1988)}]{1988ApJ...334..771V}
{van Dishoeck} E.~F., {Black} J.~H., 1988, \apj, 334, 771

\bibitem[{Wang {et~al}\mbox{.}(2014)Wang, Qiu, Shi, Zhang, Fang, \&
  Zhang}]{27050d31659a4cb1bdc9a59f0dde3f51}
Wang J., Qiu J., Shi Y., Zhang J., Fang M., Zhang Z., 2014, Astrophysical
  Journal, 796

\bibitem[{{Wang} {et~al}\mbox{.}(2009){Wang}, {Chin}, {Henkel}, {Whiteoak}, \&
  {Cunningham}}]{2009ApJ...690..580W}
{Wang} M., {Chin} Y.-N., {Henkel} C., {Whiteoak} J.~B., {Cunningham} M., 2009,
  \apj, 690, 580

\bibitem[{{Wang} {et~al}\mbox{.}(2004){Wang}, {Henkel}, {Chin}, {Whiteoak},
  {Hunt Cunningham}, {Mauersberger}, \& {Muders}}]{2004A&A...422..883W}
{Wang} M., {Henkel} C., {Chin} Y.-N., {Whiteoak} J.~B., {Hunt Cunningham} M.,
  {Mauersberger} R., {Muders} D., 2004, \aap, 422, 883

\bibitem[{{Watanabe} {et~al}\mbox{.}(2014){Watanabe}, {Sakai}, {Sorai}, \&
  {Yamamoto}}]{2014ApJ...788....4W}
{Watanabe} Y., {Sakai} N., {Sorai} K., {Yamamoto} S., 2014, \apj, 788, 4

\bibitem[{{Wilson} \& {Rood}(1994)}]{1994ARA&A..32..191W}
{Wilson} T.~L., {Rood} R., 1994, \araa, 32, 191

\bibitem[{{Wu} {et~al}\mbox{.}(2005){Wu}, {Evans}, {Gao}, {Solomon}, {Shirley},
  \& {Vanden Bout}}]{2005ApJ...635L.173W}
{Wu} J., {Evans}, II N.~J., {Gao} Y., {Solomon} P.~M., {Shirley} Y.~L., {Vanden
  Bout} P.~A., 2005, \apjl, 635, L173

\end{thebibliography}

\bsp	
\label{lastpage}
\end{document}